# Spinorial Reduction of the Superdimensional Dual-covariant Field Theory[*]

Yaroslav Derbenev
derbenev@jlab.org

*Thomas Jefferson National Accelerator Facility, Newport News, VA 23606, USA*

Abstract

In this paper we produce further specification of the geometric and algebraic properties of the earlier introduced *superdimensional dual-covariant field theory* (SFT) in a *N*-dimensional *unified manifold* (UM) [1] as an approach to a *unified field theory* (UFT). Considerations in the present paper are directed by a requirement of *transformational invariance* (TI) of connections of derivatives of *dual state vector* (DSV) and *unified gauge field* (UGF matrices) to these objects themselves established by mean of *N split metric* matrices of a rank $\mu$ (SM, an extended analog of Dirac matrices) in frame of the related Euler-Lagrange (EL) equations for DSV and UGF derived in [1]. TI requirement is posed on SFT as one of the aspects of the general demand of *irreducibility* claimed to UFT; it leads to *rotational invariance* of SM and *grand metric* tensor (GM) as being structured on SM. Study in this work has led to explication of geometrical nature of SM and DSV as *spin-affinors* (variable in space of the *unified manifold*) and dual *spin-field*, respectively, in accordance with the E. Cartan's *theory of spinors* [2 - 4], while UGF fields are recognized as *boson* objects. Algebraic EL equations on SM are derived applying method of the *Lagrange multipliers* to take into account *spinorial reduction* of SM.

## 1. Introduction

A *superdimensional dual-covariant field theory* (SFT) as an approach to unified field theory (UFT) has been exposed earlier in paper [1]. System of differential equations for *dual state vector* (DSV) field $\boldsymbol{\Psi} \equiv \Psi^\alpha$, $\boldsymbol{\Phi} \equiv \Phi_\alpha$ and related *coefficient functions* (CF): *split metric* (SM) matrices $\boldsymbol{\Lambda}^k \equiv \Lambda^{\alpha k}_\beta$ and *unified gauge field* (UGF) matrices $\boldsymbol{\mathcal{A}}_k \equiv \mathcal{A}^{\alpha k}_\beta$ ($k = 1, 2, \ldots N;\ \ \alpha, \beta = 1, 2, \ldots \mu$) as geometrical objects in *N*- dimensional *unified manifold* (UM) of variables $\breve{\psi}^k$ was derived based on system of principles unified by the demand of the *irreducibility*. The developed treat, however, left open a question about connection of transformations of DSV to transformations of the UM variables. The present paper is called to fill in this gap by adding a requirement of *transformational invariance* principle (TI) of SFT equations.

A concise review of the SFT principles and equations for DSV and UGF treated earlier in [1] are presented Chapters 2 and 3. Adding TI to the system of SFT principles leads directly to TI of SM matrices and *grand metric* (GM) tensor as structured on SM. Analysis of this imposition impact to SFT produced in Chapters 4 through 6 leads to explication of geometrical nature of SM as *spin-matrices* and DSV as *spinor (*or *Fermi) field*, respectively, according to Cartan's *theory of spinors* in an even $N$ dimensions space [2 - 4]. Correspondently, UGF is recognized as a system of the *boson* type fields. Cartan's *commutation rules* for SM are then introduced to the SFT Lagrangian applying method of *Lagrange multipliers*. In result, EL equations on DSV and UGF do not change compared to those of paper [1], while equations on SM experience a *spinorial reduction*.

## 2. Basic principles

*1. SFT as a self-contained field theory.* The SFT is profiled as a *self-contained* differential theory for State Vector field (in the further texts denoted $\Xi$) as function of variables of a *unified manifold* (UM), subordinate of a *differential law* (DL). This definition, in accordance to the SV status, implies that, in such theory derivatives of SV on UM variables are connected to SV itself – via *coefficient functions* (CF) as objects of the theory which are not given (proposed) in advance as explicit functions of UM variables but connected to SV by the related differential equations. Note that, Maxwell-Dirac electrodynamics as a field theory of the *first level* (i.e. a "classical" field theory) is a theory of this type. Let us use notation $\check{\psi}$ to denote a *point* in UM given by $N$ "coordinates" $\check{\psi}^k$:

$$\check{\psi} \equiv \{\check{\psi}^k\}; \quad k = 1, 2, \ldots, N. \tag{2.1}$$

In these notations, the transition from theory of classical fields $\hat{\psi}(\hat{x})$, $\hat{x} = \vec{r}, t$ to a unified theory can be symbolized in the following manner:

$$\hat{\psi}(\hat{x}) \to \Xi(\check{\psi}); \quad \check{\psi} \equiv (\hat{\psi}, \hat{x}); \quad \frac{\partial \hat{\psi}(\hat{x})}{\partial \hat{x}} \to \frac{\partial}{\partial \check{\psi}} \Xi(\check{\psi}) \tag{2.2}$$

In the context of the relation to space of variables of QFT, UM of SFT should be envisioned as space of degrees of freedom corresponding (but not being an identity) to the *fermion* fields of QFT, according to the point of view of the author that fermion's degrees of freedom should play a pilot role in a fundamental irreducible theory of the micro-world. In more direct comparison, fermion features of the theory should be addressed to transformation properties of the State Vector field as the global pilot object of SFT. *Boson* objects (fields) can be envisioned to be profiled based on binary combinations of the fermion type of objects. In principle, transformation properties of the *observable objects* should result from the UFT dynamics. In our sight, a background soil for appearance of bosons should be associated with such coefficient functions of equation for SV as *gauge objects* $\check{\mathcal{A}}(\check{\psi})$ introduced for covariant extension of the SV derivatives [1]:

$$\frac{\partial}{\partial \check{\psi}} \Xi \to \left(\frac{\partial}{\partial \check{\psi}} + \check{\mathcal{A}}\right) \Xi. \tag{2.3}$$

It may seem at the first glance that SFT cannot lead to *quantum* properties of field dynamics in projection to the 4-dimensional space-time manifold, unless we incorporate postulates of

---

[1] Standard Model of QFT includes *Higgs boson* as a non-gauge field serving for creation of particles' masses. Also, the *supersymmetry* theories suggest unification of fermions and bosons (including *gravitons*) in one extended group of objects. In the context of the developments towards UFT, we prefer to keep a point of view expressed by the above comment.

"quantization" in the concept. Remind in this connection that, the non-relativistic wave mechanics of Schrödinger with "non-quantized" wave function leads immediately to quantization of atomic energy levels and *uncertainty relations* concerning the transition to classical mechanics, as well as the relativistic theory of Dirac immediately explains electron spin and introduces a concept of creation-annihilation of particles before the "secondary quantization". After all, QFT (initiated by P. Dirac) as a mathematical system in essence is a *superdimensional* differential theory for *state vector* (SV) of a dynamical system as function of a conglomerate of variables (space-time and variety of *fields* as free variables associated with "elementary particles"), subordinate to a *Schrödinger equation* with *energy operator* ("quantum Hamiltonian") of a certain structure which includes differentials over all variables. In this context, the presented treat is in general correspondence to the QFT establishment. Description of *matter* in terms of the State Vector field in approach to UFT as a self-contained field theory in space of Unified Manifold might be implied corresponding to quest for a *Universal Wave Function* by S. Hawking [1].

*2. Unified Manifold as space of free numbers.* The c*oordinates* i.e. *variables* of the Unified Manifold of SFT should be regarded as *free numbers* varying continuously. They cannot be referred to "material bodies", "classical objects", etc. Such references are not compatible with the sense of UFT as a fundamental, *irreducible* field theory.

*3. Dynamical genesis of the physical geometry.* No specific geometrical characterization of UM space should be posed in advance. Definitions of *distance* or *interval* should not be introduced in advance, as well. Geometrical characteristics (metric signature, topology of UM, group properties of SFT objects, etc.) can only be profiled based on the established structure and *solutions* of the SFT differential system.

*4. The Unified Manifold− Matter Function homomorphism.* In order to be irreducible, the fundamental differential law should be associated with a procedure of a background level produced on the UM space, which could be a special *homomorphism* of $N$ variables $\check{\psi}^k$ i.e. existence of $\mu$ functions $\varphi^\alpha(\check{\psi})$ directed by a *differential law* (DL), subject to find out:

$$\check{\psi}^k \rightarrow \varphi^\alpha(\check{\psi}); \quad k = 1,2,\ldots,N; \quad \alpha = 1,2,\ldots,\mu \tag{2.4}$$

we will call this homomorphism *matter function* (MF). DL should be associated with derivatives of MF that define a homomorphic connection between the differentials:

$$d\varphi^\alpha = \frac{\partial \varphi^\alpha}{\partial \check{\psi}^k} d\check{\psi}^k ; \tag{2.5}$$

in a symbolic form:

$$d\boldsymbol{\varphi} = F d\check{\boldsymbol{\psi}}; \qquad F \equiv F_k^\alpha \equiv \frac{\partial \varphi^\alpha}{\partial \check{\psi}^k} . \tag{2.6}$$

We will talk about system of derivatives $F_k^\alpha$ as *homomorphic matrix*, meaning that, generally, it is not quadratic. Note that, the *inverse connection* cannot be formulated similar to equation (2.5), once $\mu > N$.

*5. Dynamic connection between UM and MF transformations.* At transformation of UM variables at a point with matrix $A$, MF differentials are transformed with some matrix $B$:

$$d\breve{\psi} \rightarrow d\breve{\psi}' = A d\breve{\psi}; \qquad d\varphi \rightarrow d\varphi' = B d\varphi = BF d\breve{\psi}. \qquad (2.7)$$

$$d\varphi' = F' d\breve{\psi}' = F' A d\breve{\psi}; \qquad \rightarrow \qquad F' = BFA^{-1} \qquad (2.8)$$

*6. State Vector as field of directions in MF space.* Object of DL, *state vector* field (SV) is supposed to be associated with differentials of MF. DL, however, cannot be derived immediately for functions $\varphi^\alpha$; but it could be derived for a *system of functions collinear* with differentials $d\varphi^\alpha$ of MF, denote them $\Psi^\alpha$:

$$\Psi^\alpha \propto d\varphi^\alpha; \quad \rightarrow \quad \Psi^\alpha = \frac{d\varphi^\alpha}{d\tau} = \frac{\partial \varphi^\alpha}{\partial \breve{\psi}^k} \frac{d\breve{\psi}^k}{d\tau} = F_k^\alpha \mathfrak{q}^k; \quad \mathfrak{q}^k = \frac{d\breve{\psi}^k}{d\tau}; \quad \boldsymbol{\Psi} = F\mathbf{q}. \qquad (2.9)$$

Here we used notations $\boldsymbol{\Psi}$ and $\mathbf{q}$ for $\Psi^\alpha$ and $\mathfrak{q}^k$ as for (contravariant) vectors referred to the MF and UM spaces, respectively. Object $\Psi^\alpha$ thus can be associated with *tangent vectors* of "world lines" $\varphi^\alpha(\tau)$ in MF space corresponding to world lines $\breve{\psi}^k(\tau)$ in UM space ($\tau$ is a *canonical parameter* of a line [1]). Considering transformations of UM variables, we can write:

$$\mathbf{q} \rightarrow \mathbf{q}' = A\mathbf{q}; \qquad \boldsymbol{\Psi} \rightarrow \boldsymbol{\Psi}' = B\boldsymbol{\Psi}. \qquad (2.10)$$

Geometrical nature of *state vector* field $\Psi^\alpha$ with respect to the unified manifold, i.e. its transformation at transformation of UM variables should be determined by the *differential law*, subject to find out.

*7. Affine duality of SFT.* In parallel with field $\boldsymbol{\Psi}$, one can consider a $\mu$-components vector field independent of $\boldsymbol{\Psi}$ but associated with the *inverse* transformation in MF space; denote it $\Phi_\alpha$ or $\boldsymbol{\Phi}$ (*covariant state vector* field, CSV):

$$\boldsymbol{\Phi}' = \boldsymbol{\Phi} B^{-1}; \qquad (2.11)$$

If in matrix terms SV can be represented by a column of numbers (functions), then $\boldsymbol{\Phi}$ is represented by a row of numbers. One also may consider an inverse homomorphism between covariant vectors in MF and UM space as follows:

$$p_k = F_k^\alpha \Phi_\alpha. \qquad (2.12)$$

*8. State norm.* Scalar product of the two introduced vector objects is *invariant* of transformations:

$$\mathbb{N} \equiv (\boldsymbol{\Phi \Psi}) \equiv \Phi_\alpha \Psi^\alpha = \mathbb{N}' = invariant; \qquad (2.13)$$

the identical invariance takes place in the UM space, since:

$$\Phi_\alpha \Psi^\alpha = \Phi_\alpha F_k^\alpha \mathfrak{q}^k = p_k \mathfrak{q}^k. \qquad (2.14)$$

We will consider not just one vector field but *dual couple* of *real* vector fields as a *master object* of a unified field theory:

$$\Xi \Rightarrow (\Psi^\alpha, \Phi_\alpha). \tag{2.15}$$

We call the introduced duality the *affine duality,* and association of these two vector fields the *Dual State Vector* field (DSV). Note that, this duality distinguishes essentially from the conventional *metric duality* usually obtained by lifting up or down of indices of the objects applying *metric tensor* like in GTR: two the above introduced vector fields are considered as two systems of $N$ numbers (functions of the UM variables) *independent* of each other. Also note that, field $\Phi_\alpha$ is not associated with gradients of a *scalar field* but is of a more general nature. After all, in our view the scalar objects of irreducible theory should not be introduced to theory (i.e. postulated) to play role of the basic objects but could only be *composed* as invariant forms based on use of the *affine duality* and (or) *metric tensor* of the theory. Our *affine duality* also distinguishes from the *complex numbers* duality of QFT.

Introduction of a covariant state vector field $\mathbf{\Phi}$ as dual but independent partner to contravariant field $\mathbf{\Psi}$ is a start point for building up the SFT *dynamics*.

*9. Duality of SV as a presage of UM – MF algebraic structural isomorphism.* Duality of SV can be viewed as a prerequisite for building the structural (i.e. *tensor*) forms in MF space based on DSV and its derivatives as functions of UM variables. This sight leads to consideration of possible *structural* or *algebraic UM – MF isomorphism.*

*10. The homogeneity principle.* Differential system of SFT is considered *homogeneous* in space of UM. This implies that, the differential system should be formulated only as relations between involved basic objects $X^a$ and their derivatives $\partial_k X^a$ and should not include any explicit i.e. *given* functions of UM variables. This requirement is one of those that make SFT a *self-contained* theory.

The sense of this principle consists in the following. Establishing the differential law as relations between SV and its derivatives takes introduction of *coefficient functions*. When profiling these relations, one should use no assumptions about behavior of these objects as functions of the manifold variables, neither *ad hoc* or with references to "reality". Instead, coefficient functions should be connected to SV by the correspondent differential equations, as above mentioned.

This principle may seem a "routine" one at first glance, since it is a basic declaration of QFT as a *quantum* field theory in the 4-dimensional space-time manifold. On the other hand and *in fact*, QFT is a differential field theory for SV as function of variables of $N > 4$ dimensions manifold. However, QFT does not follow the homogeneity principle when building up the dynamic law for the SV as the *secondary quantization* function: while considering SV *in fact* as function of fields $Q$, QFT at the same time utilizes representation of *Schrödinger equation* in which Hamiltonian is an *explicit given function* (form) of $Q$. This methodology cannot bring QFT to the class of the self-contained theories.

*11. The uniformity principle.* Equations of SFT should be *uniform* (symmetric, *homogenized*) over all components of the involved objects and UM variables.

***12. Dynamical genesis of the transformation properties.*** Transformation properties (hence, *"geometrical nature"*) not only of DSV but all SFT objects should be viewed as determined by the SFT dynamic laws.

***13. Differential irreducibility of SFT as principle of a dynamical existence.*** Equations for DSV should connect first derivative of this object to itself and not include higher order derivatives. EL equations on the triadic objects (coefficient functions) should be formulated in the lowest order derivatives, in correspondence to the all *irreducibility* demands.

This demand can be considered as expression of an ontological principle of the *dynamical existence*. Namely, DL of DSV is an autonomic system of $2\mu$ equations in first order derivatives for $2\mu$ functions of UM variables. DSV as an object subordinate to such a law does not have points of zero: if there would be one such point, then DSV would be zero everywhere. Search for the correspondence $N \rightarrow \mu$, in our sight, should be directed by principle of an *irreducible structural UM-MF isomorphism* in cooperation with all the posed *irreducibility* demands. Exploration for establishing of dimensionality $N$ goes beyond the frame of this publication.

***14. Preview of master equations.*** Similar to QFT, differential equations for DSV are meant to have a form linear on DSV. The law should connect the *first derivatives* of DSV to DSV itself, in accordance with the irreducibility requirements. So it necessarily includes some multi-index connection objects, the *coefficient functions* (CF) matrices. As a primer attempt, one could write:

$$P_\beta^{\alpha k} \partial_k \Psi^\beta = \Psi^\alpha ; \qquad \overline{P}_\alpha^{\beta k} \partial_k \Phi_\beta = \Phi_\alpha . \tag{2.16}$$

***15. Constraint of covariance***

Primer form of DSV equations (2.16), however, does not meet requirement of homogeneity because of appearance of terms with derivatives of matrix $B$ at transformations, − so equations (2.16) are not *covariant*. This defect can supposedly be fixed by introduction of compensated terms in equations (2.16) in the following way:

$$\partial_k \Psi^\beta \rightarrow \mathfrak{D}_k \Psi^\beta \equiv \partial_k \Psi^\beta + \mathcal{A}_{\gamma k}^\beta \Psi^\gamma ; \qquad \partial_k \Phi_\beta \rightarrow \mathfrak{D}_k \Phi_\beta \equiv \partial_k \Phi_\beta - \mathcal{A}_{\beta k}^\gamma \Phi_\gamma \tag{2.17}$$

Each of the 3-indices *geometrical objects*: $P_\beta^{\alpha k}$; $\overline{P}_\alpha^{\beta k}$ and $\mathcal{A}_{\beta k}^\alpha$ is association of $N$ matrices (on Greek indices) of rank $\mu$. Introducing notations: $\mathcal{A}_{\beta k}^\alpha \equiv \mathcal{A}$; $P_\beta^{\alpha k} \equiv \mathbf{P}$; $\overline{P}_\alpha^{\beta k} \equiv \overline{\mathbf{P}}$, we can write equations (2.16) in the following symbolic view:

$$\mathbf{P} \cdot \mathfrak{D}\mathbf{\Psi} = \mathbf{\Psi}; \qquad \mathfrak{D}\mathbf{\Phi} \cdot \overline{\mathbf{P}} = \mathbf{\Phi}; \tag{2.18}$$

$$\mathfrak{D}\mathbf{\Psi} \equiv (\partial + \mathcal{A})\mathbf{\Psi} ; \qquad \mathfrak{D}\mathbf{\Phi} \equiv \partial \mathbf{\Phi} - \mathbf{\Phi} \cdot \mathcal{A} ; \tag{2.19}$$

Requirement of compensation of derivatives of matrix $B$ at trtnasormations leads to the following transformation law for object $\mathcal{A}_{\beta k}^\alpha$ [1]:

$$\mathcal{A} \rightarrow \mathcal{A}' = A^{-1} B (\mathcal{A} + \partial) B^{-1} . \tag{2.20}$$

We call matrices $\mathcal{A}$ *unified gauge field* (UGF) in the context of an external correspondence to gauge fields of QFT. In the context of the correspondence to objects of the conventional differential geometry, they can be characterized as *hybrid affine tensor* or *hybrid Christoffel symbols*, being transformed with two different matrices, $B$ and $A$. Objects $\mathfrak{D}_k\boldsymbol{\Psi}$ and $\mathfrak{D}_k\boldsymbol{\Phi}$ are called *covariant deri*vatives of DSV.

*16. **Reality of the SFT objects as an attribute of general covariance***. The *Differential Law* (DL) as an analytical algorithm of UFT should be formulated in all *real* terms. The *imaginary unit* "*i*", and *complex* objects or variables are not admitted.

This "puritanical" restriction is imposed due to a consideration that the presence of *invariable objects*, like "*i*", is not compatible with the requirement of *general covariance* that implies that all the involved objects should be variable in a *covariant way*. We presume that, the complex analytical structure of the existing "quantum theory" shall be recognized in frame of SFT as a particular asymptotical sector of a more general analytical structure of SFT represented in terms of such background objects as vectors and matrices of UM and (or) MF space − all real.

*17. **Hybrid tensors**.* Once UGF matrices are introduced as object transformed according to equation (2.20), then objects (2.19) are transformed similar to tensors but with two different matrices, $A$ and $B$:

$$\mathfrak{D}'\boldsymbol{\Psi}' = A^{-1}B\mathfrak{D}\boldsymbol{\Psi}; \qquad \mathfrak{D}'\boldsymbol{\Phi}' = A^{-1}(\mathfrak{D}\boldsymbol{\Phi})B^{-1}. \qquad (2.21)$$

Further, based on presumed equations (2.18), transformation properties (2.21) lead to transformation rule for triadic objects $\mathbf{P}$ and $\overline{\mathbf{P}}$ as follows:

$$\mathbf{P} \to \mathbf{P}' = AB\mathbf{P}B^{-1}; \qquad \overline{\mathbf{P}} \to \overline{\mathbf{P}}' = AB\overline{\mathbf{P}}B^{-1}. \qquad (2.22)$$

Objects $\mathfrak{D}\boldsymbol{\Psi}, \mathfrak{D}\boldsymbol{\Phi}$ and $\boldsymbol{P}, \overline{\boldsymbol{P}}$ can be regarded as *hybrid tensors* (or *h-tensors*). Matrices $\mathbf{P}$ and $\overline{\mathbf{P}}$ can be characterized as *triadic h-tensors*.

*18. **Covariant derivatives of the hybrid tensors**.* Introduction of covariant derivatives of h-tensors requires the involvement of *Christoffel symbols* $\Gamma_{kl}^m$ (affine tensor) of differential geometry:

$$\Gamma_{lm}^k \equiv \frac{1}{2}w^{kn}(\partial_l w_{nm} + \partial_m w_{nl} - \partial_n w_{lm}), \qquad (2.23)$$

where $w_{kl}$ is an even-symmetric non-degenerated covariant tensor (metric tensor). In particular:

$$\mathfrak{D}_l t_{\alpha k} \equiv \partial_l t_{\alpha k} - \mathcal{A}_{\alpha l}^\beta t_{\beta k} - \Gamma_{kl}^m t_{\alpha m}. \qquad (2.24)$$

$$\mathfrak{D}_l t_k^\alpha \equiv \partial_l t_k^\alpha + \mathcal{A}_{\beta l}^\alpha t_k^\beta - \Gamma_{kl}^m t_m^\alpha; \qquad (2.25)$$

$$\mathfrak{D}_l t^{\alpha k} \equiv \partial_l t^{\alpha k} + \mathcal{A}_{\beta l}^\alpha t^{\beta k} + \Gamma_{ml}^k t^{\alpha m}; \qquad (2.26)$$

$$\mathfrak{D}_l t_\alpha^k \equiv \partial_l t_\alpha^k - \mathcal{A}_{\alpha l}^\beta t_\beta^k + \Gamma_{ml}^k t_\alpha^m. \tag{2.27}$$

$$\mathfrak{D}_l t_\beta^{\alpha k} \equiv \partial_l t_\beta^{\alpha k} + \mathcal{A}_{\gamma l}^\alpha t_\beta^{\gamma k} - \mathcal{A}_{\beta l}^\gamma t_\gamma^{\alpha k} + \Gamma_{ml}^k t_\beta^{\alpha m} ;$$

$$\mathfrak{D}_l t_{\beta k}^\alpha \equiv \partial_l t_{\beta k}^\alpha + \mathcal{A}_{\gamma l}^\alpha t_{\beta k}^\gamma - \mathcal{A}_{\beta l}^\gamma t_{\gamma k}^\alpha - \Gamma_{kl}^m t_{\beta m}^\alpha ; \tag{2.28}$$

$$\mathfrak{D}_l t_\beta^{\alpha km} \equiv \partial_l t_\beta^{\alpha km} + \mathcal{A}_{\gamma l}^\alpha t_\beta^{\gamma km} - \mathcal{A}_{\beta l}^\gamma t_\gamma^{\alpha km} + \Gamma_{nl}^k t_\beta^{\alpha nm} + \Gamma_{nl}^m t_\beta^{\alpha kn} , \text{ etc.} \tag{2.29}$$

Apparently, tensor $w^{kl}$ (inverse to $w_{kl}$) can be structured as binary product of the triadic h-tensors contracted on Greek indices (see below).

*19. Requirement of the existence of a conservative supercurrent.* Existence of a conservative vector current (*supercurrent*) should be one of the necessary features of the derived differential system. Such property should follow immediately from equations for DSV as an attribute of the dynamical existence principle for a dual state vector field. Conservative vector current associated with DSV can be presented in the following form:

$$\mathcal{J}^k = \Lambda_\beta^{\alpha k} \Phi_\alpha \Psi^\beta; \quad \nabla_k \mathcal{J}^k \equiv \frac{1}{\sqrt{w}} \partial_k(\sqrt{w} \mathcal{J}^k) = 0; \quad w = |det w_{kl}| \tag{2.30}$$

with undefined h-tensor $\Lambda_\beta^{\alpha k} \equiv \mathbf{\Lambda}^k$ and even-symmetric covariant tensor $w_{kl}$. Further, let us represent h-tensors $\mathbf{P}^k$ and $\overline{\mathbf{P}}^k$ in equations (2.18) in the following way, introducing an undefined s-tensor $\boldsymbol{\lambda} \equiv \lambda_\beta^\alpha$:

$$\mathbf{P}^k = (\mathbf{1} + \boldsymbol{\lambda})^{-1} \mathbf{\Lambda}^k; \qquad \overline{\mathbf{P}}^k = -\mathbf{\Lambda}^k(\mathbf{1} - \boldsymbol{\lambda})^{-1} . \tag{2.31}$$

Drawing then the requirement of a *conservative* current (2.30) and using equations (2.31), we find solution for s-tensor $2\boldsymbol{\lambda}$ as *covariant divergence* of h-tensor $\mathbf{\Lambda}^k$:

$$2\boldsymbol{\lambda} \Rightarrow \mathfrak{D}_k \mathbf{\Lambda}^k \equiv \frac{1}{\sqrt{w}} \partial_k(\sqrt{w} \mathbf{\Lambda}^k) + [\mathcal{A}_k , \mathbf{\Lambda}^k] ; \tag{2.32}$$

here symbol $[;]$ means commutator of two matrices. Thus, specification of DSV equations for existence of a conservative current (2.30) might result in the following form of these equations:

$$\mathbf{\Lambda}^k \mathfrak{D}_k \mathbf{\Psi} + \left(\frac{1}{2} \mathfrak{D}_k \mathbf{\Lambda}^k + \mathbf{1}\right) \mathbf{\Psi} = 0; \qquad (\mathfrak{D}_k \mathbf{\Phi}) \mathbf{\Lambda}^k + \mathbf{\Phi} \left(\frac{1}{2} \mathfrak{D}_k \mathbf{\Lambda}^k - \mathbf{1}\right) = 0, \tag{2.33}$$

with reduction of a couple triadic h-tensors, matrices $\mathbf{P}^k$ and $\overline{\mathbf{P}}^k$ to *one h-tensor*, N matrices $\mathbf{\Lambda}^k$.

*20. Metric tensor of UM.* Symmetric covariant tensor can be introduced not as independent relative of already introduced objects but as inverse to contravariant tensor $\Lambda^{kl}$ structured on h-tensor $\Lambda_\beta^{\alpha k}$ as follows:

$$w_{kl} \Rightarrow \Lambda_{kl}; \qquad \Lambda^{km}\Lambda_{lm} = \Delta_l^k; \qquad \Lambda^{kl} \equiv \Lambda_\beta^{\alpha k}\Lambda_\alpha^{\beta l} = \frac{1}{\mu}Tr(\mathbf{\Lambda}^k\mathbf{\Lambda}^l). \qquad (2.34)$$

Then:

$$\mathfrak{D}_k\mathbf{\Lambda}^k \Rightarrow \frac{1}{\sqrt{\Lambda}}\partial_k(\sqrt{\Lambda}\mathbf{\Lambda}^k) + [\mathcal{A}_k, \mathbf{\Lambda}^k]; \quad \Lambda = |det\Lambda_{kl}| \qquad (2.35)$$

$$\partial_k(\sqrt{\Lambda}\mathcal{J}^k) = 0. \qquad (2.36)$$

**21. Constraint of CFs – DSV coupling.** CFs $\mathbf{\Lambda}^k$ and $\mathcal{A}_k$ should not be *given* i.e. explicated in advance as functions of UM variables, in view of the *homogeneity* principle. They also cannot be viewed as constant matrices, since such foundation would be contrary to the general covariance principle. Then, there is the only path to resolution of this issue: CFs should be *connected to DSV* by other equations based on some fundamental principle of the differential calculus.
The following constraints have being directing search for the CFs – DSV coupling.
- H-tensor $\mathbf{\Lambda}^k$ might be connected DSV, its covariant derivatives and h-tensor forms which could be built on $\mathcal{A}_k$ and its derivatives – if such forms do exist.
- Gauge field $\mathcal{A}_k$ cannot be directly connected to h-tensor forms in a similar way, since it is not transformed as an h-tensor.
- Connection of UGF to DSV and SM might be realized if there exist an h-tensor form composed on UGF and its derivatives.

**22. Abandoning the superposition principle.** Once the CFs of master equations (2.33) are connected to DSV, the differential system of SFT arrives non-linear in DSV, thus violating the *superposition principle,* a basic postulate of QFT.

**23. Hybrid curvature form as covariant derivative of gauge** Let us consider the second covariant derivatives of s-fields $\Psi^\alpha$ and $\Phi_\alpha$. Considering $\mathfrak{D}_k\Psi^\alpha$ and $\mathfrak{D}_k\Phi_\alpha$ as h-tensors, we can write:

$$\mathfrak{D}_k\mathfrak{D}_l\Psi^\alpha = \partial_k\mathfrak{D}_l\Psi^\alpha + \mathcal{A}_{\beta k}^\alpha\mathfrak{D}_l\Psi^\beta - \Gamma_{lk}^m\mathfrak{D}_m\Psi^\alpha;$$

Calculating then the alternated second covariant derivatives, we find:

$$(\mathfrak{D}_k\mathfrak{D}_l - \mathfrak{D}_l\mathfrak{D}_k)\Psi^\alpha = \mathfrak{R}_{\beta kl}^\alpha\Psi^\beta;$$

similar:

$$(\mathfrak{D}_k\mathfrak{D}_l - \mathfrak{D}_l\mathfrak{D}_k)\Phi_\alpha = -\mathfrak{R}_{\alpha kl}^\beta\Phi_\beta;$$

here

$$\mathfrak{R}_{\beta kl}^\alpha \equiv \partial_k\mathcal{A}_{\beta l}^\alpha - \partial_l\mathcal{A}_{\beta k}^\alpha + \mathcal{A}_{\gamma k}^\alpha\mathcal{A}_{\beta l}^\gamma - \mathcal{A}_{\gamma l}^\alpha\mathcal{A}_{\beta k}^\gamma, \qquad (2.37)$$

or, in symbols of matrix:

$$\mathbf{\mathfrak{R}}_{kl} = \partial_k\mathcal{A}_l - \partial_l\mathcal{A}_k + [\mathcal{A}_k, \mathcal{A}_l]. \qquad (2.38)$$

We call h-tensor form $\mathfrak{R}_{kl}$ *hybrid curvature form* (HCF); it is recognized as *covariant derivative of UGF* $\mathcal{A}_k$ [1*].

**24. Geometrical objects of SFT** Treat of a DSV-based field theory requires introduction of covariant derivatives of the hybrid objects including covariant derivative of UGF itself, in the analogy to the *Riemann-Christoffel curvature form* of the differential geometry. Conventional *tensors* (including *vectors*) as objects that are transformed only with matrix *A* can be structured on the introduced basic objects and their covariant derivatives. There also can be composed the multi-Greek index objects as transformed only with matrix *B*; we call such ones the *s-tensors*. *Scalar functions* as *invariants of transformations* of UM variables and MF objects all result *in dynamics* from *scalar forms* composed *in presupposition*s of the *Extreme Action* principle.

**25. Presumed gauge equations.** A simplest connection of UGF to DSV in SFT differential system can be presumed as *matrix extension of Maxwell' equations*:

$$\mathfrak{D}_l \mathfrak{R}^{kl} = \mathcal{J}^k ; \qquad \mathfrak{R}^{kl} \equiv \Lambda^{km}\Lambda^{ln}\mathfrak{R}_{mn} \tag{2.39}$$

$$\mathcal{J}^k \equiv \frac{1}{2}(\Lambda^k \mathbf{N} + \mathbf{N}\Lambda^k); \quad \mathbf{N} \equiv \mathbf{\Phi} \times \mathbf{\Psi}; \quad Tr\mathcal{J}^k = J^k ; \tag{2.40}$$

here:

$$\mathfrak{D}_l \mathfrak{R}^{kl} \equiv \frac{1}{\sqrt{\Lambda}} \partial_l(\sqrt{\Lambda}\mathfrak{R}^{kl}) + [\mathcal{A}_l; \mathfrak{R}^{kl}] \tag{2.41}$$

is *covariant divergence of HCF*.

**26. Principle of Transformational Invariance**

G*eneral covariance* as one of the guiding principles of SFT had been directing the preceding steps of derivations. It is critically important that structure of SFT under study admits possibility to consider objects transformed with two different matrices engaged in a transformation law; yet ranks of these two matrices may distinct. It allows one to produce a reduction of the covariance principle agreeable with a physical requirement of *transformational invariance* of differential law of the SFT as dual-covariant field theory.

*TI of SFT as invariance of SM and GM at transformations*

System of equations for DSV and UGF (2.33) and (2.39) can be viewed as relations between these objects and their derivative by mean of coefficient functions which all are Split Metric matrices and determinant of Grand Metric tensor (built on SM) and their derivatives. Apparently, physical requirement of *transformational invariance of DL form* can be formulated as invariance of matrices $\mathbf{\Lambda}^k$ at transformation of UM variables (see (2.22)):

$$\mathbf{\Lambda}' \equiv AB\mathbf{\Lambda}B^{-1} = \mathbf{\Lambda}. \tag{2.42}$$

Equation (2.42) can be considered as determining matrix *B* as functional of transformation matrix *A* and Split Metric matrices $\mathbf{\Lambda}^k$:

$$B = B(A, \mathbf{\Lambda}).  \tag{2.43}$$

Note that, rank of matrix $B$ is equal to that of $\mathbf{\Lambda}^k$ on definition according to equations on DSV.

*Presumed SM correlations*
Further, it can be anticipated that, linear on matrix $B$ nature of equations (2.42) may pose specific restrictions on algebraic structure of matrices $\mathbf{\Lambda}^k$. These restrictions can manifest in a system of coupling equations for matrices $\mathbf{\Lambda}^k$ that may result from analysis of equations (2.42):

$$\mathbf{C}^{\tilde{q}}(\mathbf{\Lambda}) = 0;  \tag{2.44}$$

here $\mathbf{C}^{\tilde{q}}$ are some specific forms structured on SM matrices.

*Transformation paradigm in SFT*
- A consistent fundamental field theory should admit only transformations which meet requirement of invariance of its differential law. This "restriction" is considered one of specification aspects of the *irreducibility* demand concerning the role and position of *transformations* in a unified theory.
- *Transformation parameters* are viewed as dynamic entities of solution of the SFT differential system.
- This point of view on the *transformation paradigm* is considered as one of the ontological principles of SFT.

**27. Extreme Action as principle of a dynamic balance.** We resort to this principle in order to derive connections of the introduced triadic objects $\mathcal{A}$ and $\mathbf{\Lambda}$ to DSV, together with equations for DSV itself:

$$\delta \int \mathcal{L}(X, \partial X) d\Omega = 0; \quad \mathcal{L} \equiv \mathrm{L}\sqrt{\Lambda}; \quad d\Omega \equiv d\check{\psi}^1 d\check{\psi}^2 \ldots d\check{\psi}^N;  \tag{2.45}$$

posing, as usual, variations of basic objects $\delta X = 0$ at a (arbitrary) closed surface limiting volume of integration. *Lagrangian form* $\mathcal{L}$ is structured on basic objects $X$ and their derivatives $\partial X$ as product of *scalar Lagrangian* form $\mathrm{L}$ and *weigh factor* $\sqrt{\Lambda}$, so that $\sqrt{\Lambda} d\Omega$ is *invariant differential volume*. EAP generally results in *Euler-Lagrange* (EL) *equations* for system of basic objects :

$$\partial_k \frac{\partial \mathcal{L}}{\partial(\partial_k X^a)} = \frac{\partial \mathcal{L}}{\partial X^a}.  \tag{2.46}$$

EAP is a unique methodological principle for deriving the fundamental equations of a field theory. It is one of the corner stones of the QFT methodology, though the way it is used therein – building the "quantum Hamiltonian" (energy operator) by a transition from Lagrangian of a "classical" ("non-quantized") field theory – look more like a mnemonic rule or postulated receipt rather than a logically conditioned principle. In approach to SFT as "classical" field theory in a superdimensional manifold, the Euler-Lagrange equations (including, of course, EL equations for DSV itself) are immediately derived as a fundamental law of the theory with no resorting to further procedures as "quantization", etc. The superdimensional EAP is viewed as replacing the quantization paradigm of QFT; quantum behavior of the observable material objects could be interpreted as associated with projecting of a superdimensional field dynamics to the *intelligible* 4-dimensional space-time manifold (STM). Dimensionalities $N$ and $\mu$ of the theory are supposed to be determined in the frame of the theory itself as a minimum required for a self-consistent irreducible SFT. An associated "home task" of the theory should be explanation of special STM role as a realm for the intelligible world, immediately grasped by the senses and macro-apparatus.

## 28. Preliminary Lagrangian form

Under press of the above listed *irreducibility* demands, scalar form L and tensor $\Lambda_{kl}$ are composed on basic objects in the following way:

$$\mathrm{L} = \mathbb{L} + \mathbb{L}_{TI}; \qquad \mathbb{L} \equiv \mathbb{M} + \mathbb{G}; \qquad \mathbb{M} \equiv \mathbb{N} + \mathbb{D}; \qquad \mathbb{L}_{TI} \equiv \mathbf{M}_q \mathbf{C}^q(\Lambda); \qquad (2.47)$$

$$\mathbb{N} \equiv \Phi_\alpha \Psi^\alpha; \qquad \mathbb{D} \equiv \Lambda_\alpha^{\beta k} \mathfrak{D}_{\beta k}^\alpha; \qquad (2.48)$$

$$\mathfrak{D}_{\beta k}^\alpha \equiv \frac{1}{2}(\Phi_\beta \mathfrak{D}_k \Psi^\alpha - \Psi^\alpha \mathfrak{D}_k \Phi_\beta) \equiv \mathfrak{D}_k; \qquad (2.49)$$

$$\mathbb{G} \equiv \frac{1}{4} \Lambda^{kl} \Lambda^{mn} \mathbb{G}_{km;ln}; \qquad \mathbb{G}_{km;ln} \equiv Tr(\mathfrak{R}_{km} \mathfrak{R}_{ln}). \qquad (2.50)$$

Part $\mathbb{L}$ in composition of L is scalar Lagrangian of a theory in which transformation of objects ($\Psi^\alpha$; $\Phi_\alpha$) are not connected to transformation of the *unified manifold* variables as treated in [1]. Term $\mathbb{L}_{TI}$ is introduced following method of *Lagrange multipliers* in order to take into account connections between matrices $\Lambda_{\beta k}^\alpha$ presumed in general by equations (2.44). According to this method, Lagrange multipliers (LM) $\mathbf{M}_q$ are considered as the additional basic objects, subjects of independent variations in the *extreme action* derivations.

To remind, the Greek and the Roman indices do not interfere in structure of Lagrangian, being associated with transformations in the MF and UM space, respectively. Forms $\mathfrak{D}_k \Psi^\alpha$ and $\mathfrak{D}_k \Phi_\beta$ are *covariant derivatives* of DSV given by formulas (2.17) or (2.19). Form $\mathfrak{D}_{\beta k}^\alpha$ is named *matter matrices* (MM), and form $\mathfrak{R}_{\beta k l}^\alpha$ *hybrid curvature form*, HCF (symbol [ , ] denotes commutator of two matrices); the last one is uniquely recognized as *covariant derivative of gauge* $\mathcal{A}_{\beta k}^\alpha$ itself. Tensor form $\mathbb{G}_{km;ln}$ is named *gauge 4-tensor*. Scalar forms $\mathbb{N}$, $\mathbb{D}$, $\mathbb{M}$ and $\mathbb{G}$ are named *state norm, kinetic scalar, matter scalar* and *gauge scalar*, respectively. Note that, all definitions (2.48) through (2.50) are unambiguous, since contractions between Roman and Greek indices are not legitimate in the differential theory under treat.

*29. Scale Invariance.* Principles of building up the unified theory should eliminate sensitivity of its dynamical properties to introduction of arbitrary real constants as multipliers at scalar items of Lagrangian. We call such property *scale invariance* (SI), considering it as a feature necessary for a field theory to be a candidate in UFT. Scalar Lagrangian (2.47) as well as the whole Lagrangian (2.45) is *scale-invariant* i.e. it possesses the *immunity* of its form relative introduction of arbitrary real numbers (positive or negative) as multipliers of its scalar items: by a proper simple scaling the DSV, SM and LM magnitudes, whole the scalar Lagrangian can be returned to the initial form (2.47) [1*]. Same is true relative the introduction of arbitrary real multipliers of UGF; in this case the restoring of Lagrangian form is achieved by the correspondent re-scaling of the UM variables. It should be noted, by the way, that SI as an intrinsic property of the constant-less irreducible field theory can be implemented in a logically consistent way only based on EAP. *This commitment makes EAP an indispensable background dynamical principle, no-alternative receipt of deriving basic equations of a unified covariant differential field theory.*

SI of form $\mathbb{L}$ has been proved in paper [1]; SI of form L of the *invariant theory* treated in the present paper is then obvious.

*30. The mini-max principle.* To be in a consistence with the *irreducibility* demand, Lagrangian of UFT should be subordinate to the *mini-max principle*: while under the restrictive press of the exhibited requirements, number of different scalar forms composing the Lagrangian should be *maximum* at *minimum collection* of the basic objects. In our case, when weigh factor $\sqrt{\Lambda}$ is structured as shown above (also under press of the irreducibility demand), the scale invariance and the mini-max requirement both are referred directly to the scalar Lagrangian L. Any additions to L which could be built on the same basic objects (including LM) violate the feature of *scale invariance*.

The exhibited set of principles can be considered as specification for implementation of a general and universal requirement that should be asserted to a fundamental field theory - the *irreducibility* principle, in accordance with legacy of W. Pauli [1].

*31. Hamilton-Nöther equations* As generally known [1, 2], when considering *complete* derivatives $\partial_k \mathcal{L}$ of Lagrangian $\mathcal{L}$, one finds the following generic *dynamic identities*:

$$\partial_k \mathcal{L} = \partial_l [\frac{\partial \mathcal{L}}{\partial (\partial_l X^a)} \partial_k X^a] ; \qquad (2.51)$$

or

$$\partial_l (\sqrt{w} \mathcal{H}_k^l) = 0 ; \qquad (2.52)$$

here we introduced a mixed valence 2 *pseudo-tensor* object:

$$\mathcal{H}_k^l \equiv \frac{1}{\sqrt{w}} \frac{\partial (L\sqrt{w})}{\partial (\partial_l X^a)} \partial_k X^a - \Delta_k^l L . \qquad (2.53)$$

In our case, taking into account that, forms $w_{kl} = \Lambda_{kl}$ and $\mathbb{L}_{TI}$ do not include derivatives of basic objects, and that *in dynamics* $\mathbb{L}_{TI} = 0$, we find:

$$\mathcal{H}_k^l \equiv \frac{\partial \mathbb{L}}{\partial (\partial_l X^a)} \partial_k X^a - \Delta_k^l \mathbb{L} \qquad (2.54)$$

Performing variation derivatives, we find the following generic *dynamic identities* as equations for a mix valence 2 *pseudo-tensor* object $\mathcal{T}_k^l$:

$$\partial_l(\sqrt{\Lambda}\mathcal{J}_k^l) = 0 \,; \tag{2.55}$$

$$\mathcal{J}_k^l \equiv \frac{1}{2}\Lambda_\beta^{\alpha l}(\Phi_\alpha \partial_k \Psi^\beta - \Psi^\beta \partial_k \Phi_\alpha) + \mathfrak{R}_\beta^{\alpha l m}\partial_k \mathcal{A}_{\alpha m}^\beta + \mathbb{L}\Delta_k^l \,. \tag{2.56}$$

It should be noted that, though this object is not tensor, its structure (i.e. form as a composition of basic objects and their derivatives) does not change at arbitrary transformations of UM variables.

### 3. Euler-Lagrange equations on DSV and UGF

#### 3.1. Master equations

EL equations on DSV in the explicit matrix view have the same structure as derived in [1]:

$$\mathbf{\Lambda}^k \partial_k \mathbf{\Psi} + \frac{1}{2}(\hat{\partial}_k \mathbf{\Lambda}^k + \mathbf{\Lambda}^k \mathcal{A}_k + \mathcal{A}_k \mathbf{\Lambda}^k)\mathbf{\Psi} + \mathbf{\Psi} = 0 \,; \tag{3.1}$$

$$(\partial_k \mathbf{\Phi})\mathbf{\Lambda}^k + \frac{1}{2}\mathbf{\Phi}(\hat{\partial}_k \mathbf{\Lambda}^k - \mathbf{\Lambda}^k \mathcal{A}_k - \mathcal{A}_k \mathbf{\Lambda}^k) - \mathbf{\Phi} = 0 \,; \tag{3.2}$$

$$\hat{\partial}_k \equiv \partial_k + \frac{\partial_k \Lambda}{2\Lambda} \tag{3.3}$$

Symbols $\mathbf{\Psi}$ and $\mathbf{\Phi}$ denote *column* and *row* of components $\Psi^\alpha$ and $\Phi_\alpha$, respectively. Taking into account definition of DSV covariant derivatives (2.6), we can write EL equations on DSV in the following symbolic covariant form:

$$\mathbf{\Lambda}^k \mathfrak{D}_k \mathbf{\Psi} + (\frac{1}{2}\mathfrak{D}_k \mathbf{\Lambda}^k + 1)\mathbf{\Psi} = 0 \,; \qquad (\mathfrak{D}_k \mathbf{\Phi})\mathbf{\Lambda}^k + \mathbf{\Phi}(\frac{1}{2}\mathfrak{D}_k \mathbf{\Lambda}^k - 1) = 0 \,. \tag{3.4}$$

Here

$$\mathfrak{D}_k \mathbf{\Psi} \equiv \partial_k \mathbf{\Psi} + \mathcal{A}_k \mathbf{\Psi} \,; \qquad \mathfrak{D}_k \mathbf{\Phi} \equiv \partial_k \mathbf{\Phi} - \mathbf{\Phi}\mathcal{A}_k \,; \tag{3.5}$$

$$\mathfrak{D}_k \mathbf{\Lambda}^k \equiv \hat{\partial}_k \mathbf{\Lambda}^k + [\mathcal{A}_k, \mathbf{\Lambda}^k] \,. \tag{3.6}$$

Object $\mathfrak{D}_k \mathbf{\Lambda}^k$ is *covariant divergence* of Split Metric matrices $\Lambda_\beta^{\alpha k}$.

***Scalar dynamic identities of DSV equations***

Contracting equations (3.1) in products with $\Phi_\alpha$, while equations (3.2) in products with $\Psi^\alpha$, and taking sum and difference of the resulting scalar equations, we find the following two scalar equations [1].
*Conservation of the supercurrent:*

$$\hat{\partial}_k \mathcal{J}^k = 0 \,; \qquad \mathcal{J}^k \equiv \Psi^\alpha \Lambda_\alpha^{\beta k}\Phi_\beta \,. \tag{3.7}$$

*Nullification of matter scalar in dynamics*

Other an important direct consequence of DSV equations is nullification *in dynamics* of matter scalar form $\mathbb{M}$ i.e. there is a *dynamical identity*:

$$\mathbb{D} = -\mathbb{N}; \qquad (3.8)$$

then *in dynamics*

$$\mathbb{L} \Longrightarrow \mathbb{G}. \qquad (3.9)$$

### *Reduction of Hamilton-Noether equations*

Taking into account dynamical identity (3.9), we can make the correspondent replacement in form of pseudo-tensor (2.56):

$$\mathcal{T}_k^l \Longrightarrow \frac{1}{2}\Lambda_\beta^{\alpha l}(\Phi_\alpha \partial_k \Psi^\beta - \Psi^\beta \partial_k \Phi_\alpha) + \mathfrak{R}_\beta^{\alpha lm}\partial_k \mathcal{A}_{\alpha m}^\beta + \mathbb{G}\Delta_k^l. \qquad (3.10)$$

### 3.2. Gauge equations

EL equations on UGF also do not change compared to those derived in [1]. In an explicit matrix form they have the following view:

$$\hat{\partial}_l \boldsymbol{\mathfrak{R}}^{kl} + \boldsymbol{\mathcal{A}}_l \boldsymbol{\mathfrak{R}}^{kl} - \boldsymbol{\mathfrak{R}}^{kl}\boldsymbol{\mathcal{A}}_l = \frac{1}{2}(\boldsymbol{\Lambda}^k \mathbf{N} + \mathbf{N}\boldsymbol{\Lambda}^k); \qquad \mathbf{N} \equiv \boldsymbol{\Phi} \cdot \boldsymbol{\Psi} \qquad (3.11)$$

Here

$$\boldsymbol{\mathfrak{R}}^{kl} \equiv \Lambda^{km}\Lambda^{ln}\boldsymbol{\mathfrak{R}}_{mn}. \qquad (3.12)$$

Gauge equations can be written in a symbolic covariant form, as well:

$$\mathfrak{D}_l \boldsymbol{\mathfrak{R}}^{kl} = \boldsymbol{\mathcal{J}}^k; \qquad (3.13)$$

with *covariant divergence* of hybrid tensor $\boldsymbol{\mathfrak{R}}^{kl}$:

$$\mathfrak{D}_l \boldsymbol{\mathfrak{R}}^{kl} \equiv \hat{\partial}_l \boldsymbol{\mathfrak{R}}^{kl} + [\boldsymbol{\mathcal{A}}_l, \boldsymbol{\mathfrak{R}}^{kl}] \qquad (3.14)$$

on the left-hand side, and *supercurrent* matrix $\mathcal{J}_\beta^{\alpha k} \equiv \boldsymbol{\mathcal{J}}^k$ on the right-hand side:

$$\boldsymbol{\mathcal{J}}^k \equiv \frac{1}{2}\{\boldsymbol{\Lambda}^k, \mathbf{N}\}. \qquad (3.15)$$

Gauge equations (3.13) connect the affine h-tensor, $\mathcal{A}_{\beta k}^\alpha$ to DSV and SM.

### *Contracted gauge equations*
### *Extended Faraday equations*

By taking trace of HCF form (2.38) on Greek indices $\beta = \alpha$ we obtain a skew-symmetric covariant tensor $\mathbb{F}_{kl}$ defined as:

$$\mathbb{F}_{kl} \equiv \frac{1}{\mu}\mathfrak{R}^{\alpha}_{\alpha kl} = \partial_k \mathcal{A}_l - \partial_l \mathcal{A}_k = -\mathbb{F}_{lk}\,; \qquad \mathcal{A}_k \equiv \frac{1}{\mu}\mathcal{A}^{\alpha}_{\alpha k} \qquad (3.16)$$

Note that object $\mathbb{F}_{kl}$ is tensor despite that $\mathcal{A}^{\alpha}_{\alpha k}$ is not a vector. This tensor satisfies the identity equations similar to the *first pair* of Maxwell equations (we call them *Faraday equations*, but now in case of $N$-dimensional space of UM):

$$\partial_m \mathbb{F}_{kl} + \partial_l \mathbb{F}_{mk} + \partial_k \mathbb{F}_{lm} = 0\,. \qquad (3.17)$$

*Extended Maxwell equations*

By taking trace of gauge equations (3.13) on Greek indices $\beta = \alpha$ we obtain the following $N$ equations (similar to the *second pair* of Maxwell equations):

$$\hat{\partial}_l \mathbb{F}^{kl} = \mathcal{J}^k. \qquad (3.18)$$

These $N$ equations connect two contravariant objects: a *contravariant skew-symmetric* tensor field:

$$\mathbb{F}^{kl} \equiv \Lambda^{km}\Lambda^{ln}\mathbb{F}_{mn} = -\mathbb{F}^{lk} \qquad (3.19)$$

and a contravariant *vector* field, the *supercurrent* (2.30).

Note that, equations (2.18) are in a direct consistence with equation (3.7), since $\check{\partial}_k \check{\partial}_l \mathbb{F}^{kl} = \check{\partial}_l \check{\partial}_k \mathbb{F}^{kl} \equiv 0$, as for any skew-symmetric contravariant tensor.

## 4. Preliminary metric equations

Now we have to consider EL equations on rest of basic objects of Lagrangian (2.45): equations on SM matrices $\mathbf{\Lambda}^k$ (note that Lagrangian does not include derivatives of both):

$$\frac{\partial \mathcal{L}}{\partial \mathbf{\Lambda}^k} = 0 \qquad (4.1)$$

and equations on multipliers $\mathbf{M}_q$:

$$\frac{\partial \mathcal{L}}{\partial \mathbf{M}_q} \equiv \sqrt{\Lambda}\mathbf{C}^q(\mathbf{\Lambda}) = 0\,; \qquad (4.2)$$

the last ones are simply the connections equations (2.44). Equations on SM distinct from equations derived in [1] by terms with multipliers $M_q$ in product with derivatives of forms $\Lambda^q$ on SM:

$$(\mathbb{G}_{kl} - L\Lambda_{kl})\Lambda^l + \mathbf{M}_q \frac{\partial \mathbf{C}^q(\mathbf{\Lambda})}{\partial \mathbf{\Lambda}^k} = -\mathfrak{D}_k\,; \qquad (4.3)$$

here:

$$\mathbb{G}_{kl} \equiv \Lambda^{mn}\mathbb{G}_{km;ln}\,. \qquad (4.4)$$

Taking into account reduction of Lagrangian in dynamics $L \Rightarrow \mathbb{G}$, we can produce this replacement in equations (4.3), in result at this stage we have the following system of EL equations on matrices $\mathbf{\Lambda}^k$ and multipliers $\mathbf{M}_q$:

$$(\mathbb{G}_{kl} - \mathbb{G}\Lambda_{kl})\mathbf{\Lambda}^l + \mathbf{M}_q \frac{\partial \mathbf{C}^q(\Lambda)}{\partial \mathbf{\Lambda}^k} = -\mathfrak{D}_k \; ; \tag{4.5}$$

$$\mathbf{C}^q(\Lambda) = 0 \,. \tag{4.6}$$

Forms $\mathbf{C}^q(\Lambda)$ will be specified below after analysis of equations of SM invariance (2.42).

## 5. Transformational invariance of SFT

*Invariance of Grand Metric*
Since in the considered SFT system Grand Metric tensor $\Lambda^{kl}$ is not an independent object but structured on SM according to equation (2.34), then, consequently, GM also is invariant of transformations:

$$\Lambda^{kl} = inv \quad \rightarrow \quad \Lambda_{kl} = inv \quad \rightarrow \quad \Lambda = inv \,. \tag{5.1}$$

We call this property in association with property (2.42) *metric invariance* (MI) of SFT. Two consequences follow directly from property (5.1):
1. *Restriction on type of transformations*:

$$detA = \pm 1.$$

This follows from general equations of GM transformation:

$$\Lambda'_{kl} \equiv A_k^m A_l^n \Lambda_{mn} \; ; \tag{5.2}$$

in our case

$$\Lambda'_{kl} = \Lambda_{kl} \; ; \quad \rightarrow \quad A_k^m A_l^n \Lambda_{mn} = \Lambda_{kl} \; ; \quad \rightarrow \quad (detA)^2 = 1. \tag{5.3}$$

Transformations at $detA = 1$ are *rotations* or *proper rotations*; transformations at $detA = -1$ are *reflections* or *improper rotations*. We keep of a view that *transformations* as a category of physics have a dynamical genesis and can be described in terms of integration of infinitely small (infinitesimal) transformations for which matrix $A$ is close to unit; so we have to accept case of normal rotations:

$$detA = 1. \tag{5.4}$$

2. *Invariance of the interval form*
*Metric invariance* (5.1) leads to invariance of *form* of the generally invariant *differential bi-interval:*

$$\mathbb{I} \equiv \Lambda_{kl} d\breve{\psi}^k d\breve{\psi}^l \,. \tag{5.5}$$

Indeed, let us denote for the simplicity sake:

We then can write:
$$d\breve{\psi}^k \equiv x^k.$$

$$\mathbb{I}' = \Lambda'_{kl} x'^k x'^l = \mathbb{I} = \Lambda_{kl} x^k x^l.$$

Since $\Lambda'_{kl} = \Lambda_{kl}$, then:

$$\Lambda_{kl} x^k x^l = \Lambda_{kl} x'^k x'^l, \tag{5.6}$$

i.e. *form* of bi-interval (5.5) does not change at the *metric-invariant* transformations.

## 6. Invariant reduction of Split Metric

### 6.1. Equations of infinitesimal transformations

At first, we will specify MI equations (5.1) and (5.4) for *infinitesimal* transformations. *Finite* transformation can be explicated by integration of the infinitesimal ones.

*General connection $B(A)$ for infinitesimal transformations*

It will be a little more convenient to re-write equation (2.42) in the followingan equivalent view:

$$B^{-1} \Lambda B = A \Lambda \tag{6.1}$$

and explicate it relative the Roman indices:

$$B^{-1} \Lambda^n B = A^n_k \Lambda^k. \tag{6.2}$$

For *infinitesimal* transformation matrices $A$ and $B$ can be represented as follows:

$$A \Rightarrow \tilde{A} \equiv 1 + a \equiv \Delta^n_k + a^n_k; \quad \Delta^n_k = \begin{cases} 0, & n \neq k \\ 1, & n = k \end{cases}; \quad a \equiv a^n_k \Rightarrow 0; \tag{6.3}$$

$$B \Rightarrow \tilde{B} = 1 + b; \qquad b \Rightarrow 0. \tag{6.4}$$

Neglecting then the second order terms in equations:

$$(1 - b) \Lambda^n (1 + b) = (\Delta^n_k + a^n_k) \Lambda^k, \tag{6.5}$$

we obtain equations for relations between infinitesimal matrices $b$ and $a^n_k$:

$$\Lambda^n b - b \Lambda^n = \Lambda^k a^n_k. \tag{6.6}$$

*Invariant infinitesimal transformations of the UM space*

Before considering equations for invariance of SM matrices $\mathbf{\Lambda}^k$, let us look at MI equations (5.4) for Grand Metric tensor $\Lambda_{kl}$. Now we write them according to representation (6.3):

$$(\Delta_k^m + a_k^m)(\Delta_l^n + a_l^n)\Lambda_{mn} = \Lambda_{kl}. \tag{6.7}$$

For the infinitesimal transformations we find the following relations posed on matrix $a_k^l$ by the MI requirement:

$$(a_l^n \Delta_k^m + a_k^m \Delta_l^n)\Lambda_{mn} = a_l^n \Lambda_{kn} + a_k^m \Lambda_{lm} = 0. \tag{6.8}$$

Introducing notations

$$a_{kl} \equiv a_k^m \Lambda_{lm}, \tag{6.9}$$

we find the following conditions:

$$a_{kl} + a_{lk} = 0$$

i.e.:

$$a_{kl} = -a_{lk}; \tag{6.10}$$

or

$$a_k^m \Lambda_{lm} = -a_l^m \Lambda_{km}. \tag{6.11}$$

Also, inversing relations (6.9):

$$a_k^l = \Lambda^{lm} a_{km}, \tag{6.12}$$

we find:

$$Tra \equiv a_k^k = \Lambda^{kl} a_{kl} = 0. \tag{6.13}$$

*Metric-invariant infinitesimal transformations as rotations*

Two vectors $x^k$ and $y^k$ are called *orthogonal*, if their *scalar product* is zero:

$$\Lambda_{kl} x^k y^l = 0. \tag{6.14}$$

Considering invariance of bi-interval (5.5) at infinitesimal transformation:

$$\delta(\Lambda_{kl} x^k x^l) = 2\Lambda_{kl} x^k \delta x^l = 0; \tag{6.15}$$

so in this case $y^l \equiv \delta x^l$; we find that at infinitesimal transformation preserving the bi-interval form (5.5) variation of vector $x^k$ is orthogonal to the vector. Such transformations are called *rotations*.

*Reduction of MI equations for SM taking into account the rotation symmetry*

Let us now re-write equations (6.6) in terms of matrix $a_{kl}$ on the right-hand side using definition (6.9) and taking into account the skew symmetry of $a_{kl}$ :

$$\boldsymbol{\Lambda}^n b - b\boldsymbol{\Lambda}^n = \frac{1}{2}(\Lambda^{kn}\boldsymbol{\Lambda}^l - \Lambda^{ln}\boldsymbol{\Lambda}^k)a_{kl}. \tag{6.16}$$

Since both $a_{kl}$ and $b$ are considered infinitesimal, we can re-present $b$ via a system of finite matrices (on Greek indices of a rank $\mu$) $\boldsymbol{\sigma}^{kl}$ in the following way :

$$b = \frac{1}{4}\boldsymbol{\sigma}^{kl}a_{kl}; \qquad \boldsymbol{\sigma}^{kl} = -\boldsymbol{\sigma}^{lk}. \tag{6.17}$$

Since matrix $a_{kl}$ commutes on definition with $b$, equations (6.16) can be re-written in the following view:

$$(\boldsymbol{\Lambda}^n\boldsymbol{\sigma}^{kl} - \boldsymbol{\sigma}^{kl}\boldsymbol{\Lambda}^n)a_{kl} = 2(\Lambda^{kn}\boldsymbol{\Lambda}^l - \Lambda^{ln}\boldsymbol{\Lambda}^k)a_{kl}. \tag{6.18}$$

Matrix $a_{kl}$ is skew-symmetric on its indices $k, l$ but is arbitrary in rest, therefore we can equalize coefficient matrices (on Greek indices) between left and right sides of equations (6.18), so we obtain equations to define the finite matrices $\boldsymbol{\sigma}^{kl}$:

$$\boldsymbol{\Lambda}^n\boldsymbol{\sigma}^{kl} - \boldsymbol{\sigma}^{kl}\boldsymbol{\Lambda}^n = 2(\Lambda^{kn}\boldsymbol{\Lambda}^l - \Lambda^{ln}\boldsymbol{\Lambda}^k). \tag{6.19}$$

### 6.2. Solution for matrix of the infinitesimal rotations

#### 6.2.1. General solution
According to equations (6.19), matrices $\boldsymbol{\sigma}^{kl}$ can be considered as analytical functions of SM (i.e. they should be composed on SM) which do not commute with SM. Taking into account their skew-symmetry on Roman indices, one can see the only possible representation for these objects:

$$\boldsymbol{\sigma}^{kl} \Rightarrow \beta \cdot (\boldsymbol{\Lambda}^k\boldsymbol{\Lambda}^l - \boldsymbol{\Lambda}^l\boldsymbol{\Lambda}^k). \tag{6.20}$$

where β is a scalar form or number. Note that, matrices $\boldsymbol{\sigma}^{kl}$ and $b$ possess the following features:

$$Tr\boldsymbol{\sigma}^{kl} = 0; \quad \rightarrow \quad Trb = \frac{1}{4}a_{kl}Tr\boldsymbol{\sigma}^{kl} = 0. \tag{6.21}$$

Substituting form (6.20) in equations (6.19) results in the following system of algebraic equations as conditions posed on SM matrices:

$$\boldsymbol{\Lambda}^n(\boldsymbol{\Lambda}^k\boldsymbol{\Lambda}^l - \boldsymbol{\Lambda}^l\boldsymbol{\Lambda}^k) - (\boldsymbol{\Lambda}^k\boldsymbol{\Lambda}^l - \boldsymbol{\Lambda}^l\boldsymbol{\Lambda}^k)\boldsymbol{\Lambda}^n = \frac{2}{\beta}(\Lambda^{kn}\boldsymbol{\Lambda}^l - \Lambda^{ln}\boldsymbol{\Lambda}^k), \tag{6.22}$$

which can be rewritten in the following view:

$$\{\{\boldsymbol{\Lambda}^n, \boldsymbol{\Lambda}^k\}, \boldsymbol{\Lambda}^l\} - \{\{\boldsymbol{\Lambda}^n, \boldsymbol{\Lambda}^l\}, \boldsymbol{\Lambda}^k\} = \frac{2}{\beta}(\Lambda^{kn}\boldsymbol{\Lambda}^l - \Lambda^{ln}\boldsymbol{\Lambda}^k); \tag{6.23}$$

here brackets $\{,\}$ denote anti-commutator of two matrices; in particular:

$$\{\mathbf{\Lambda}^k, \mathbf{\Lambda}^l\} \equiv \mathbf{\Lambda}^k\mathbf{\Lambda}^l + \mathbf{\Lambda}^l\mathbf{\Lambda}^k.$$

Finally, let us rewrite equations (6.23) in the following view:

$$\{(\beta\{\mathbf{\Lambda}^k, \mathbf{\Lambda}^n\} - \Lambda^{kn} \cdot \mathbf{1}), \mathbf{\Lambda}^l\} - \{(\beta\{\mathbf{\Lambda}^l, \mathbf{\Lambda}^n\} - \Lambda^{ln} \cdot \mathbf{1}), \mathbf{\Lambda}^k\} = 0. \tag{6.24}$$

Apparently, these equations suggest that anti-commutator of SM matrices should be a proportion of unit matrix (on Greek indices) i.e.:

$$\beta\{\mathbf{\Lambda}^k, \mathbf{\Lambda}^n\} = \Lambda^{kn} \cdot \mathbf{1}. \tag{6.25}$$

Taking trace of these equations and using definition of Grand Metric $\Lambda^{kl}$ (2.34):

$$\Lambda^{kl} \equiv \frac{1}{\mu} Tr(\mathbf{\Lambda}^k \mathbf{\Lambda}^l),$$

we find that scalar $\beta$ is simply a number:

$$\beta = \frac{1}{2}. \tag{6.26}$$

So we then find solution for matrices $\boldsymbol{\sigma}^{kl}$:

$$\boldsymbol{\sigma}^{kl} \Rightarrow \frac{1}{2}(\mathbf{\Lambda}^k\mathbf{\Lambda}^l - \mathbf{\Lambda}^l\mathbf{\Lambda}^k) \tag{6.27}$$

under conditions that SM matrices $\mathbf{\Lambda}^k$ satisfy system of algebraic equations as follows:

$$\mathbf{\Lambda}^k\mathbf{\Lambda}^l + \mathbf{\Lambda}^l\mathbf{\Lambda}^k = 2\Lambda^{kl} \cdot \mathbf{1}; \tag{6.28}$$

in particular,

$$(\mathbf{\Lambda}^k)^2 = \Lambda^{kk} \cdot \mathbf{1}. \tag{6.29}$$

Problem of solution for matrices $\boldsymbol{\sigma}^{kl}$ as analytical functions of SM is thus reduced to finding SM as system of $N$ matrices of rank $\mu$ satisfying equations (6.28).

### 6.2.2. Orthogonal reduction
Tensor $\Lambda^{kl}$ can be turned to a diagonal one:

$$\Lambda^{kl} = \begin{cases} 0, & k \neq l \\ \Lambda^{kk}, & k = l \end{cases}. \tag{6.30}$$

by transformation of $N$ directions at a point of UM to an *orthogonal basis* (on definition!). Requirements (6.28) to SM matrices then will be reduced to the following ones:

$$\mathbf{\Lambda}^k\mathbf{\Lambda}^l + \mathbf{\Lambda}^l\mathbf{\Lambda}^k = \begin{cases} 0; & l \neq k \\ 2\Lambda^{kk} \cdot \mathbf{1}; & l = k \end{cases} \tag{6.31}$$

Correspondingly, such system of SM matrices can be regard as the *orthogonal* one. Solution for matrices $\sigma^{kl}$ and $b$ at orthogonal reduction then takes a short form:

$$\sigma^{kl} = \Lambda^k \Lambda^l = -\sigma^{lk}; \qquad (k \neq l). \tag{6.32}$$

System of matrices (6.31) can be normalized:

$$\Lambda^k \rightarrow \Lambda^k \sqrt{|\Lambda_{kk}|}, \tag{6.33}$$

though this normalization is only of a local mean. Equations (6.31) then take the following view:

$$\Lambda^k \Lambda^l + \Lambda^l \Lambda^k = \begin{cases} 0; & l \neq k \\ 2 sign \Lambda^{kk} \cdot \mathbf{1}; & l = k. \end{cases} \tag{6.34}$$

### 6.3. Algebra of Split Metric

#### 6.3.1. ISM-based system of linear-independent affinors

Association of *N split metric* (SM) matrices $\Lambda^k$ subordinated to commutation rules (6.31) generates algebra of affinors i.e. all orders products of SM matrices including SM themselves and also unit matrix (the last one is associated with property $(\Lambda^k)^2 = \Lambda^{kk} \cdot \mathbf{1}$):

$$\Lambda^k; \quad \Lambda^k \Lambda^l; \quad \Lambda^k \Lambda^l \Lambda^m; \ldots \quad \Lambda^1 \Lambda^2 \Lambda^3 \ldots \Lambda^N; \tag{6.35}$$

or, in general symbolic notations for these products:

$$\Lambda_q^{(k)} \equiv \Lambda^{k_{(1)}} \Lambda^{k_{(2)}} \ldots \Lambda^{k_{(q)}}; \qquad q = 1, 2, \ldots N \tag{6.36}$$

Note that, matrices $\Lambda_q^{(k)}$ are irreducible in structure and linear-independent, though basic affinors $\Lambda^k$ can be subjected by an arbitrary unitary transformation:

$$\Lambda^k \rightarrow U \Lambda^k U^{-1}. \tag{6.37}$$

We will call matrices $\Lambda_q^{(k)}$ *spin-affinors*. Matrices $\Lambda^k$ can be considered as *basic spin-affinors*.

*Total number of independent spin-affinors*
Total number of affinors $\Lambda_q^{(k)}$ (including the unit matrix) is equal to

$$\mathcal{N} = \sum_{q=0}^{N} C_q^N = 2^N. \tag{6.38}$$

Sum (6.38) results from consideration of the *Newton binomial*:

$$(a+b)^N = \sum_{q=0}^{N} C_q^N a^q b^{N-q}$$

taken at $a = b = 1$:

$$(1+1)^N = \sum_{q=0}^{N} C_q^N = 2^N.$$

### *6.3.2. Connection of UM – MF dimensionalities*

Following the irreducibility demand, collection of matrices $\Lambda_q^{(k)}$ can be viewed as basis for expansion of arbitrary matrix of a rank $\mu$ as linear composition of matrices $\Lambda_q^{(k)}$. But total number of such matrices is equal to $\mu^2$, then we can write:

$$2^N = \mu^2; \quad \rightarrow \quad \mu = 2^{\frac{N}{2}}; \tag{6.39}$$

so $N$ must be *even*:

$$N = 2\nu; \quad \mu = 2^\nu; \quad \nu = 1, 2, 3, 4, \dots . \tag{6.40}$$

In cases $\nu = 1$ and $\nu = 2$ ranks $N$ and $\mu$ coinside: $N = \mu = 2$ (Pauli matrices) and $N = \mu = 4$ (Dirac matrices), respectively. In cases $\nu = 3, 4, \dots$ we obtain $N = 6, 8, \dots$ ; $\mu = 8, 16$, etc.

We leave issue of the expansion technique, as well as explicit specification of SM beyond the scope of this paper.

### *6.3.3. Structural isomorphism of ISM algebra with Clifford algebra of poly-vectors of UM*

ISM algebra is known as *spinorial representation* of *Clifford algebra of poly-vectors* [2, 3].

*Clifford Algebra in general formalism*

In a generalized formalism, this algebra considers $N$ basic objects $\mathfrak{A}^k$ (hyper-complex numbers) of properties:

$$\mathfrak{A}^k \mathfrak{A}^l + \mathfrak{A}^l \mathfrak{A}^k = 0, \quad l \neq k ; \quad (\mathfrak{A}^k)^2 = \pm 1; \tag{6.41}$$

then there are $\mathcal{N} = 2^N$ irreducible independent products $\mathfrak{A}_q^{(k)}$:

$$\mathfrak{A}_q^{(k)} \equiv \mathfrak{A}^{k_{(1)}} \mathfrak{A}^{k_{(2)}} \dots \mathfrak{A}^{k_{(q)}}; \quad q = 1, 2, \dots N. \tag{6.42}$$

*Clifford algebra of poly-vectors of UM*

ISM algebra is isomorphic to *Clifford algebra of poly-vectors $P_q^{(k)}$* (CAP) of UM i.e. tensors composed as completely skew-symmetrized all-orders products of components $N$ basic vectors $\boldsymbol{e}_l \equiv e_l^k$ [3]:

$$P_q^{(k)} = e_{l_1}^{[k_1} e_{l_2}^{k_2} \dots e_{l_q}^{k_q]} = det(e_l^k)_q ; \quad q = 0, 1, \dots N \tag{6.43}$$

here brackets [...] denote complete alteration of indices. At *orthogonal* definition of basic vectors $\boldsymbol{e}_k$ poly-vectors $P_q^{(k)}$ can be written as:

$$P_q^{(k)} = \boldsymbol{e}_{k_1} \boldsymbol{e}_{k_2} \dots \boldsymbol{e}_{k_q} \tag{6.44}$$

meaning the property:

$$\boldsymbol{e}_k\boldsymbol{e}_l + \boldsymbol{e}_l\boldsymbol{e}_k = 0, \quad l \neq k; \quad (\boldsymbol{e}^k)^2 = \pm\mathbf{1}. \tag{6.45}$$

We call the CAP - ISM isomorphism of SFT the *structural isomorphism*.

## 7. EL equations on Split Metric

*Reduction of Lagrangian*
As in preceding paper [1], we consider SM matrices connected to DSV and UGF in frame of the extreme action principle. Now, however, EAP as variation principle has to take into account conditions (6.28) or (6.31) posed on SM by the requirement of transformational invariance of SFT equations. This can be executed in techniques of the *Lagrange multipliers* as profiled preliminary by equations (2.47). Now we can specify forms $\mathbf{C}^q(\boldsymbol{\Lambda})$ and term $\mathbb{L}_{TI}$:

$$\mathbf{C}^q(\boldsymbol{\Lambda}) \Rightarrow \mathbf{C}^{kl} \equiv \boldsymbol{\Lambda}^k\boldsymbol{\Lambda}^l + \boldsymbol{\Lambda}^l\boldsymbol{\Lambda}^k - 2\Lambda^{kl}\cdot\mathbf{1}; \tag{7.1}$$

$$\mathbb{L}_{TI} \Rightarrow \frac{1}{2}Tr(\mathbf{M}_{kl}\mathbf{C}^{kl}); \tag{7.2}$$

here matrices $\mathbf{M}_{kl} \equiv \mathrm{M}^\alpha_{\beta kl}$ are the Lagrange multipliers considered together with SM matrices $\boldsymbol{\Lambda}^k$ as basic objects, subjects of independent variations in EAP. Note that, $Tr\mathbf{C}^{kl} \equiv 0$.

*SM equations*
Using preliminary derived EL equations on SM (4.5), after performing variation derivatives of Lagrangian term (7.2) on matrices $\Lambda^{\alpha k}_\beta$ we obtain the following EL equations:

$$(\mathbb{G}_{kl} - \mathbb{G}\Lambda_{kl})\boldsymbol{\Lambda}^l + \mathbf{M}_{kl}\boldsymbol{\Lambda}^l + \boldsymbol{\Lambda}^l\mathbf{M}_{kl} - 2\boldsymbol{\Lambda}^l Tr\mathbf{M}_{kl} = -\mathfrak{D}_k. \tag{7.3}$$

EL equations on $\mathbf{M}_{kl}$:

$$\frac{\partial\mathcal{L}}{\partial\mathbf{M}_{kl}} = 0$$

simply manifest in Cartan's spin equations (6.28):

$$\boldsymbol{\Lambda}^k\boldsymbol{\Lambda}^l + \boldsymbol{\Lambda}^l\boldsymbol{\Lambda}^k = 2\Lambda^{kl}\cdot\mathbf{1}. \tag{7.4}$$

Since Lagrangian (2.45) does not include derivatives of SM, EL equations on SM result in system of *algebraic equations* on SM; so SM (consequently, GM $\Lambda_{kl}$ as well) can be considered as an *immediate* i.e. *locally defined* function of DSV, UGF and their derivatives.

*Gauge scalar as a dynamic proportion to State Norm*
Multiplying equations (7.3) by $\Lambda^{\beta k}_\alpha$, producing contraction on all indices and taking into account dynamic identity (3.8), we find the following dynamic relation:

$$(N-4)\mathbb{G} = \mathbb{D} = -\mathbb{N}. \tag{7.5}$$

So at $N \neq 4$ we find that gauge scalar $\mathbb{G}$ in the dynamics is a proportion to *state norm* $\mathbb{N}$:

$$\mathbb{G} = -\mathbb{N}/(N-4). \tag{7.6}$$

At $N \neq 4$ scalar $\mathbb{G}$ can be replaced in SM equations (7.3) and also in Hamilton- Noether pseudo-tensor (3.10) by its *dynamic identity* according to relation (7.6).

When considering case $N = 4$ in equation (7.5), we have to accept dynamic condition $\mathbb{D} = -\mathbb{N} = 0$, instead of the proportion between $\mathbb{G}$ and $\mathbb{N}$ at $N \neq 4$. It should be noted, however, that such condition for *mathematical consistence* of the theory as $\mathbb{N} = 0$ at $N = 4$ is not in complete consistence with the foundation of the *autonomic duality* of *state vector* as represented by the two *independent* vector fields in the *matter function* space, contravariant $\Psi^\alpha$ and covariant $\Phi_\alpha$. Therefore, this case should be regard as *special* in the frame of the treated superdimensional dual-covariant field theory.

*Algorithm of solution for Split Metric and GM as functions of DSV and UGF*
Equations (7.3) can be solved relative matrices $\mathbf{\Lambda}^k$ as functions of DSV, UGF, their derivatives, GM and matrices $\mathbf{M}_{kl}$. Substituting this solution in spin equations (7.4), one obtains equations for matrices $\mathbf{M}_{kl}$ as functions DSV, UGF, their derivatives and GM. In this way SM can be found as explicit functions of DSV, UGF, their derivatives and GM. Further, based on definition of GM as structured on SM according to equations (2.34), this algorithm results in system of non-linear algebraic equations for GM tensor $\Lambda_{kl}$. It should be noted that, by use of orthogonal form of Cartan's equations for SM (6.31), these equations can be reduced to equations for a diagonalized GM tensor. Finally, knowing GM, one finds SM as function of DSV, UGF and their derivatives.

### 8. Finite rotations: DSV as dual Cartan's spinor

#### 8.1. Finite rotations in Unified Manifold
Let us consider *finite* rotation in a plane $(x^k, x^l)$. Matrix of such rotation can be characterized by a single parameter $\xi$:

$$X = A(\xi)X_0. \tag{8.1}$$

Knowing matrix of infinitesimal rotation $a = a_l^k$, we can derive differential equation for matrix of finite rotation $A(\xi)$. We can write (see (6.3)):

$$(A + dA)X_0 = \tilde{A}AX_0 = (1 + a)AX_0; \quad \rightarrow \quad dA = aA. \tag{8.2}$$

Let us introduce differential $d\xi$ of parameter $\xi$ in the following way:

$$a_{(kl)} = -\sqrt{|\Lambda_{kk}\Lambda_{ll}|}d\xi; \tag{8.3}$$

then we can write:

$$a = a_l^k \equiv \boldsymbol{\alpha}d\xi; \quad \boldsymbol{\alpha} \equiv \sqrt{|\Lambda_{kk}\Lambda_{ll}|}\begin{pmatrix} 0 & \Lambda^{kk} \\ -\Lambda^{ll} & 0 \end{pmatrix}. \tag{8.4}$$

Matrix $\boldsymbol{\alpha}$ is normalized in a way that:

$$\boldsymbol{\alpha}^2 = \frac{\Lambda_{kk}\Lambda_{ll}}{|\Lambda_{kk}\Lambda_{ll}|} = \pm 1 = \begin{cases} -1; & \Lambda_{kk}\Lambda_{ll} > 0 \\ 1; & \Lambda_{kk}\Lambda_{ll} < 0 \end{cases} \tag{8.5}$$

Returning to equation (8.2), we find differential equation for matrix $A$:

$$\frac{dA}{d\xi} = \boldsymbol{\alpha} \cdot A \tag{8.6}$$

Taking into account "initial condition"

$$A(0) = 1, \tag{8.7}$$

solution of equation (8.6) is

$$A(\xi) = \exp(\boldsymbol{\alpha}\xi) \equiv \sum_{n=0}^{\infty} \frac{1}{n!} (\boldsymbol{\alpha}\xi)^n. \tag{8.8}$$

*Invariance of the transformation determinant*

Equation (8.6) satisfies requirement (5.4). Indeed, using general background formula for derivative of determinant of a matrix:

$$\frac{d}{d\xi} detA = (detA) \cdot Tr\left(\frac{dA}{d\xi} A^{-1}\right), \tag{8.9}$$

find in our case:

$$\frac{d}{d\xi} detA = (detA) Tr\boldsymbol{\alpha} = 0; \quad \rightarrow \quad detA = const = 1. \tag{8.10}$$

*Explication of rotation in a plane of UM*

Formula (8.8) can be reduced to a final view using property (8.5) of matrix . For this, let us represent expansion (8.8) in the following view:

$$A = \sum_{n=0}^{\infty} \frac{(\boldsymbol{\alpha}\xi)^{2n}}{(2n)!} + \boldsymbol{\alpha}\xi \sum_{n=0}^{\infty} \frac{(\boldsymbol{\alpha}\xi)^{2n}}{(2n+1)!}. \tag{8.11}$$

Further explication can be performed separate for two cases.
1. $\Lambda_{kk}\Lambda_{ll} > 0$.
Then:

$$\boldsymbol{\alpha} \Rightarrow \boldsymbol{\beta} \equiv \begin{pmatrix} 0; & \sqrt{\frac{\Lambda_{ll}}{\Lambda_{kk}}} \\ -\sqrt{\frac{\Lambda_{kk}}{\Lambda_{ll}}}; & 0 \end{pmatrix}; \quad \boldsymbol{\beta}^2 = -1 \tag{8.12}$$

$$A = cos\xi + \boldsymbol{\beta}sin\xi = \begin{pmatrix} cos\xi; & \sqrt{\dfrac{\Lambda_{ll}}{\Lambda_{kk}}}sin\xi \\ -\sqrt{\dfrac{\Lambda_{kk}}{\Lambda_{ll}}}sin\xi; & cos\xi \end{pmatrix}; \qquad (8.13)$$

here parameter $\xi$ can be associated with *angle* $\varphi$ of a *usual* rotation:

$$\xi \Longrightarrow \varphi; \qquad (8.14)$$

$$A(\varphi) \Longrightarrow cos\varphi + \boldsymbol{\beta}sin\varphi = \begin{pmatrix} cos\varphi; & \sqrt{\dfrac{\Lambda_{ll}}{\Lambda_{kk}}}sin\varphi \\ -\sqrt{\dfrac{\Lambda_{kk}}{\Lambda_{ll}}}sin\varphi; & cos\varphi \end{pmatrix} \qquad (8.15)$$

2. $\Lambda_{kk}\Lambda_{ll} < 0$.
Then:

$$\boldsymbol{\alpha} \Longrightarrow \boldsymbol{\gamma} \equiv \begin{pmatrix} 0; & \sqrt{-\dfrac{\Lambda_{ll}}{\Lambda_{kk}}} \\ \sqrt{-\dfrac{\Lambda_{kk}}{\Lambda_{ll}}}; & 0 \end{pmatrix}; \qquad \boldsymbol{\gamma}^2 = \mathbf{1} \qquad (8.16)$$

$$A = ch\varphi + \boldsymbol{\gamma}sh\varphi = \begin{pmatrix} ch\varphi; & \sqrt{-\dfrac{\Lambda_{ll}}{\Lambda_{kk}}}sh\varphi \\ \sqrt{-\dfrac{\Lambda_{kk}}{\Lambda_{ll}}}sh\varphi; & ch\varphi \end{pmatrix}. \qquad (8.17)$$

### *8.2. Finite rotations in Matter Function space*

Considering the correspondent transformation in MF space, we start with deriving the differential equation for the related matrix $B$. Here we can write:

$$(B + dB)\boldsymbol{\Psi}_0 = \tilde{B}B\boldsymbol{\Psi}_0 = (1 + b)B\boldsymbol{\Psi}_0; \quad \rightarrow \quad dB = bB. \qquad (8.18)$$

Using general definition of infinitesimal matrix $b$ (6.17) and reduction of matrix $a$ for rotation in a picked plane $(k, l)$ of UM according to equations (8.3) and (8.14), we obtain the following representation for correspondent matrix $b$ of DSV transformation:

$$b = \boldsymbol{\sigma}^{kl}a_{kl} = \frac{1}{2}\boldsymbol{\Lambda}^k\boldsymbol{\Lambda}^l \cdot \sqrt{|\Lambda_{kk}\Lambda_{ll}|}d\varphi \equiv \frac{1}{2}\boldsymbol{\sigma}^{(kl)}d\varphi; \qquad (8.19)$$

$$\boldsymbol{\sigma}^{(kl)} \equiv \boldsymbol{\Lambda}^k \boldsymbol{\Lambda}^l \cdot \sqrt{|\Lambda_{kk}\Lambda_{ll}|} \tag{8.20}$$

Referring back to equations (8.18), we obtain differential equation for matrix $B$:

$$\frac{dB}{d\varphi} = \frac{1}{2}\boldsymbol{\sigma}^{(kl)}B . \tag{8.21}$$

*Invariance of the transformation determinant*
Invariance of $detB$ follows immediately from equation (8.20):

$$\frac{d}{d\varphi} det B = det B \cdot Tr\left(\frac{dB}{d\varphi} \cdot B^{-1}\right) = \frac{1}{4} detB \cdot a_{kl} Tr\boldsymbol{\sigma}^{kl} = 0 ;$$

so we find:

$$det B = const = 1. \tag{8.22}$$

Apparently, this property is extendable to arbitrary *invariant transformations* that leave SM (and GM) unchanged. Such transformations can be characterized as *rotations in MF space*.

*Explication of DSV transformation at a single plane rotation in UM*
Solution for matrix $B(\varphi)$ is found in general form as follows:

$$B^{(kl)} = \exp\left(\frac{1}{2}\boldsymbol{\sigma}^{(kl)}\varphi\right) = \sum_{n=0}^{\infty}\frac{1}{n!}(\frac{1}{2}\boldsymbol{\sigma}^{(kl)}\varphi)^n. \tag{8.23}$$

Using property of matrix $\boldsymbol{\sigma}^{(kl)}$:

$$(\boldsymbol{\sigma}^{(kl)})^2 = \boldsymbol{\Lambda}^k\boldsymbol{\Lambda}^l\boldsymbol{\Lambda}^k\boldsymbol{\Lambda}^l|\Lambda_{kk}\Lambda_{ll}| = -(\boldsymbol{\Lambda}^k)^2(\boldsymbol{\Lambda}^l)^2|\Lambda_{kk}\Lambda_{ll}| = \pm\mathbf{1} \tag{8.24}$$

we can represent expansion (8.3) in explicit view as follows:

$$B^{(kl)} = \sum_{n=0}^{\infty}\frac{(\pm 1)^n}{(2n)!}(\frac{\varphi}{2})^{2n} + \boldsymbol{\sigma}^{(kl)}\sum_{n=0}^{\infty}\frac{(\pm 1)^n}{(2n+1)!}(\frac{\varphi}{2})^{2n+1}, \tag{8.25}$$

or

$$B^{(kl)} = \cos\frac{\varphi}{2} + \boldsymbol{\sigma}^{(kl)}\sin\frac{\varphi}{2} ; \qquad \Lambda_{kk}\Lambda_{ll} > 0 \tag{8.26}$$

$$B^{(kl)} = ch\frac{\varphi}{2} + \boldsymbol{\sigma}^{(kl)}sh\frac{\varphi}{2} ; \qquad \Lambda_{kk}\Lambda_{ll} < 0 . \tag{8.27}$$

### *DSV as a dual Cartan's spinor field*

Multi-component objects (systems of numbers or functions) transformed with matrices (8.26), (8.27) are known as *spinors*; they together with affinors $\boldsymbol{\Lambda}^k$ were discovered and analyzed by E. Cartan [2,3,4] as objects of analytical geometry and incepted in quantum mechanics by W. Pauli (case $\mu = 2$, non-relativistic theory) and P. Dirac (case $\mu = 4$, relativistic quantum mechanics of electrons). Their intrinsic property is uncertainty of sign due to that the sign changes after total rotation of a UM frame in $2\pi$. Objects of this type correspond to *fermions* (particles of a *half-integer spin*) according to terminology of the *elementary particle physics*. Our DSV field thus can be characterized as a *superdimensional dual fermion field*.

## 9. Transformations of Gauge

Transformation of unified gauge field (UGF) can be found directly applying general law (2.20):

$$\boldsymbol{\mathcal{A}}'_{k'} = A^k_{k'} B(\boldsymbol{\mathcal{A}}_k + \partial_k) B^{-1} = A^k_{k'} B \boldsymbol{\mathcal{A}}_k B^{-1} - B^{-1} \partial_{k'} B \qquad (9.1)$$

Consideration in detail of UGF transformations goes beyond the scope of this paper. Here we briefly touch the following aspects.
1. Matrix $B$ varies with ISM in the UM space, thus giving contribution to the term with derivatives of $B$ in transformation law (9.1). Assuming that rotation parameters $\varphi$ can be chosen constant in space, we then find:

$$\partial_{k'} B = (\partial_{k'} \boldsymbol{\sigma}^{(kl)}) \cdot \begin{cases} \sin\dfrac{\varphi}{2} \ ; & \Lambda_{mm}\Lambda_{nn} > 0 \\ sh\dfrac{\varphi}{2} \ ; & \Lambda_{mm}\Lambda_{nn} < 0 \end{cases} \qquad (9.2)$$

Terms with derivatives of rotation parameters $\varphi$ arrive if there is a reason or necessity to consider their variation in space. It is worth to note that, matrix $B$ is still inhomogeneous in UM space due to that of Split Metric – even at constant (homogeneous) rotation parameters $\varphi$. On the other hand, one might raise a question, would it be actually possible to reduce the parameterization issue, in principle, to introduction of parameters $\varphi$ constant in space. Answer this question should be a subject of further exploration of the sense the *transformation paradigm* in a covariant field theory.
2. Regardless of this issue, transformation of UGF (9.1) as being given by binary products of elements of matrix $B$ is expressed in result in terms of unambiguous analytical coefficient functions of rotation parameter $\varphi$ of UM ($sin\varphi, cos\varphi; sh\varphi, ch\varphi$), in contrary to DSV transformations given by the first power of matrix $B$ (8.26), (8.27). So UGF can be viewed corresponding to the *boson* class of QFT objects i.e. "elementary particles" of an *integer spin*.

## 10. Covariance and Invariance of SFT

The derived system of EL equations of SFT is reduced to differential equations for DSV and UGF and algebraic equations for Split Metric matrices and Grand Metric tensor as functions of DSV and UGF. All equations are *generally covariant*. Here we have to emphasize some important aspects of this feature in the context of its utilization in SFT.
1. SFT equations have been derived based on the *Extreme Action* principle (EAP). No assumptions about transformation properties (TPs) of objects and (or) about invariance of the additive items of the Lagrangian have been used when deriving the EL equations. This circumstance is not specifically associated with the presented approach to UFT but, in fact, is a routine of EAP as a conventional method of deriving fundamental equations of a field theory. In fact and in essence, TPs of basic objects and related structural forms can be found based on the derived EL equations; before i.e. when formatting the Lagrangian it is *only a presumption*. Then, TPs of multi-index objects can be explicated based on TPs of the vector type objects as fields, – once such objects have been introduced in the theory to play a master role in the differential system.

2. Being a covariant differential theory, SFT is treated in general by methods of differential geometry with its notions of vectors, tensors, affine tensors and their covariant derivatives; the last ones are defined and arrive as objects of a tensor type, as well. It should be underlined that, state vector field (DSV) associated with directions at points of the *matter function manifold* $\breve{\varphi}^\alpha$ is playing a fundamental, background role in the theory. Based on concept of a *homomorphism* $\varphi^\alpha(\breve{\psi})$, it was presumed that transformation matrix $B$ of $\mu$-components vector field $\Psi^\alpha(\breve{\psi})$ is different from matrix $A$ of transformation of UM variables $\breve{\psi}^k$: $d\breve{\psi}' = A d\breve{\psi}$; while $d\breve{\varphi}' = B d\breve{\varphi} \rightarrow \Psi' = B\Psi$.

3. Requirement of compensation for derivatives of matrix $B$ at transformation of equation on $\Psi$ leads to the following transformation law for *unified gauge field*:

$$\mathcal{A}' = A^{-1} B (\mathcal{A} + \partial) B^{-1}. \qquad (10.1)$$

4. Considering transformation of covariant vector (in terms of the MF space) field $\Phi$, assume that it transforms with a matrix $C$:

$$\Phi' = \Phi C ; \qquad (10.2)$$

Then, requirement of compensation for derivatives of matrix $C$ in equation on $\Phi$ leads to equation of transformation of gauge similar to (10.1):

$$\mathcal{A}' = A^{-1} C^{-1} (\mathcal{A} + \partial) C , \qquad (10.3)$$

so we conclude:

$$C \Rightarrow B^{-1}; \quad \rightarrow \quad \Phi' = \Phi B^{-1}. \qquad (10.4)$$

5. Now let us consider transformation of the following terms:

$$\mathfrak{D}_k \Psi \equiv \mathfrak{D}\Psi ; \qquad \mathfrak{D}_k \Phi \equiv \mathfrak{D}\Phi . \qquad (10.5)$$

Taking into account transformation law for gauge (10.1), we find:

$$\mathfrak{D}'\Psi' = A^{-1} B \mathfrak{D}\Psi ; \qquad \mathfrak{D}'\Phi' = A^{-1} (\mathfrak{D}\Phi) B^{-1} \qquad (10.6)$$

6. Returning to EL equations on DSV (3.4), we then find transformation law of matrices $\Lambda^k \equiv \Lambda$:

$$\Lambda' \equiv A B \Lambda B^{-1}. \qquad (10.7)$$

As one can see, terms (10.5) are transformed on type of tensors but with two presumably different matrices $A$ and $B$. Then collection of matrices $\Lambda^k \equiv \Lambda^{\alpha k}_\beta$ is transformed also with two kinds of matrices. We call this type of objects the *hybrid tensors*. Collection of gauge matrices $\mathcal{A}_k \equiv \mathcal{A}^\alpha_{\beta k}$ then can be characterized as the *hybrid affine tensor* or *hybrid Christoffel symbols.* Note that, in our consideration rank of the Greek (matrix) indices can be different from rank of Roman indices genetically associated with differentiation on UM variables.

7. Consequently, for term:

$$\mathfrak{D}_l \Lambda^k \equiv \partial_l \Lambda^k + \Gamma^k_{ml} \Lambda^m + [\mathcal{A}_l, \Lambda^k] \equiv \mathfrak{D} \cdot \Lambda \tag{10.8}$$

and term

$$\mathfrak{D}_k \Lambda^k \equiv \nabla_l \Lambda^k + [\mathcal{A}_k, \Lambda^k] \equiv \mathfrak{D}\Lambda \tag{10.9}$$

in equations for DSV (3.4) we find the following transformation rules:

$$(\mathfrak{D} \cdot \Lambda)' = A^{-1} B (\mathfrak{D} \cdot \Lambda) B^{-1} A, \tag{10.10}$$

$$(\mathfrak{D}\Lambda)' = B(\mathfrak{D}\Lambda) B^{-1}. \tag{10.11}$$

Apparently, object (10.8) is an h-tensor, while object (10.9) (*covariant divergence* of Split Metric) is an *s-tensor* [1].

8. Now, let us consider connection between transformation properties of the objects revealed by EL equations on gauge (3.11) or (3.13). These equations establish connection between *hybrid curvature form* (HCF) $\mathfrak{R}_{kl} \equiv \mathfrak{R}^\alpha_{\beta kl}$ (recognized as *covariant derivative of UGF*) and *supercurrent matrices* $\mathcal{J}^{\alpha k}_\beta \equiv \mathcal{J}^k$. As established in [1], form (2.39) is transformed as an h-tensor of total valence 4 with two matrices on Roman indices and two matrices on Greek indices:

$$\mathfrak{R}'_{k'l'} = A^k_{k'} A^l_{l'} B \mathfrak{R}_{kl} B^{-1} \tag{10.12}$$

if gauge $\mathcal{A}_k$ is transformed according to equation (10.1), and vice versa. The supercurrent matrices $\mathcal{J}^k$ should transform similar to the SM matrices.

9. EL equations on SM (7.3) reveal connection of SM to DSV and UGF completely agreeable with transformation properties profiled based on EL equations on DSV and UGF. System of EL equations on DSV and UGF determines transformation properties of gauge matrices $\mathcal{A}_k$ according to equation (10.1), state co-vector field $\Phi$ according to equation (10.2) and Split Metric matrices $\Lambda^k$ according to equation (10.7) – once one *assumes* that the contravariant state vector field $\Psi$ is transformed with a matrix $B$. Being an algebraic system of relations with respect to SM, EL equations on SM do not add any more specification to transformation properties of DSV, UGF and related structural forms.

10. By the way, treat of SFT in [1] left open question about connection of DSV transformations (matrix $B$) to transformations of UM variables (matrix $A$). Principle of covariance itself is not able to specify relations between transformation matrices $B$ of DSV and $A$ of UM variables in order to determine geometrical nature of DSV, SM and UGF.

11. As it was presumed in paper [1], this question finds a resolution in the context of the dynamical genesis of notion of *transformations*, namely based on the requirement of *transformational invariance* (TI) of form of SFT equations as one of the *irreducibility* demands posed on the superdimensional theory. In fact, TI is of more background meaning than even requirement of *covariance*. TI covers the commitments of the *covariance* those described above – since introduction of the *unified gauge fields* (UGF) $\mathcal{A}^\alpha_{\beta k}$ in the pre-viewed equations for DSV as a measure to compensate the derivatives of the transformation matrix $B$ actually is a primer attribute of treat of field theory with TI. But subsequent refining of features of the derived covariant uniform equations for DSV and UGF, namely posing the requirement of TI on whole differential system leads directly to *rotational* invariance of SM, consequently, to status of SM and DSV as *spin-affinors* and *spin-vector field* of E. Cartan *theory of spinors*, respectively.

12. Reduction of class of transformations of UM variables to *rotations* should be viewed not as not a voluntary action but produced under the press of the *irreducibility* demand which does cleansing the theory from uncertainty in the essential properties of the dynamical system of SFT. Further, specification of commutation properties of SM matrices to the *orthogonal* representation (6.31) according to E. Cartan [2] is in a complete correspondence with his treat of Riemannian geometry in *orthogonal frame* [5]. Moreover, with our definition of metric tensor GM of irreducible SFT as structured on SM according to equations (2.34), equations (6.31) could be viewed as suggesting a dynamical interpretation of notion of the *orthogonality* as a category of the *physical geometry* based on an *irreducible super-spinorial field theory*.

## 11. Resume

*Summary*

A purpose of this paper was profiling geometrical nature of basic objects of earlier described the *superdimensiona dual-covariant field theory* (SFT) [1]: *dual state vector* (DSV) field $\Psi^\alpha, \Phi_\alpha$, *unified gauge field* (UGF) matrices $\mathcal{A}^\alpha_{\beta k}$ and *split metric* (SM) matrices $\Lambda^{\alpha k}_\beta$. This is achieved by implementation of principle of the *transformational invariance* (TI) of the SFT differential system. The presented treatment could be regard as further going implantation of the irreducibility principle in structure of approach to UFT under investigation.

TI demand posed on the SFT concept has led to the following results.

- Utilization of TI of SM inquires specification of SM as *spin-affinors* according to *Cartan equations* (6.31).
- DSV, master object of SFT, as *geometrical object* of UM then acquires status of a *dual Fermi-field field* of a rank $\mu = 2^{(N/2)}$ ($N$ is even) transformed with matrices *B* (8.26), (8.27) for *spinors.*
- Lagrangian of SFT is modified by taking into account connection between SM i.e. commutation rules for SM as Cartan's spin-matrices.
- Earlier derived Euler-Lagrange (EL) equations on DSV and UGF do not change
- EL equations on SM are modified been complemented by terms with *Lagrange multipliers* (LM); those can be found using the Cartan's equations for SM (6.31).
- System of equations for SM and LM can be reduced to equations for *grand metric* (GM) tensor as function of tensors built on DSV, UGF and their derivatives.

*Outlook*

Here we will touch some insights and questions arising in connection to the produced spinorial reduction of SFT.

1. Based on the established theory of spinors as geometrical objects in $N = 2\nu$-dimensional space [2, 3, 4], one can presume the existence of system of basic affinors $\Lambda^k$ ($k = 1,2,...N$) as matrices of rank $2^\nu$ subordinate to conditions (6.31).

2. As noted by E. Cartan [2], spinorial theory of relativistic electrons by P. Dirac is directly extendable to a $2\nu$ - dimensions space of a Riemannian geometry based on treatment of spinors in terms of a local orthogonal frame [2]. This observation is in consistence with our profiling of metric tensor of a superdimensional Riemannian space as structured on *split metric* arriving as Cartan's affinors after their *orthogonal spinorial reduction*. By the way, orthogonal reduction of SM according to Cartan's theory of spinors is in consistence with the irreducibility principle of structuring a unified field theory.

3. On the other hand, in the presented SFT affinors $\Lambda_\beta^{\alpha k}$ are directly coupled to matter matrices $\mathfrak{D}_{\beta k}^\alpha$ (2.49) and hybrid curvature tensor $\mathfrak{R}_{\beta kl}^\alpha$ (2.37) by algebraic equations (7.3), (7.4). A question is, would the spinorial reduction of SM (which is, to be reminded, not an intricate fashion but is due to the requirement of transformational invariance of the SFT differential system) be in consistence with this coupling? The answer, strictly speaking, requires a dedicated investigation of the derived differential-algebraic system of Euler-Lagrange equations. By the way, it is worthy to note that, the dualistic nature of state vector (DSV) looks possessing a potential for an explicit implementation of spinorial properties in structure of solutions of the differential system of the theory.

4. In this paper, transformations of DSV and UGF at finite rotations have been illustrated assuming that rotation of frame of *unified manifold* (UM) is characterized by a constant parameter (rotation angle $\varphi$) despite the regional aspects of the transformations have not been treated yet. A question here is: could such simple parameterization still possible for use and consistent in Riemannian space of a self-contained *superspinorial field theory*? Investigation of this question however, goes beyond the scope of this paper.

5. Besides many critical questions to SFT in the context of its relevance to the physics of the elementary particles, one of the most intriguing questions, again, is derivability of Newton' and relativistic gravitation as an asymptotic macro-phenomenon of a superdimensional field theory in projection to the intelligible 4-dimensional space-time world.